% Our sources have nothing to hide.  The aspens are quaking.  Judy
% Miller, I want you desperately. 
%
%\let\includefigures=\iftrue
%
% the following is to use blackboard bold fonts --
\let\useblackboard=\iftrue
%
% activate this if you don't have them.
%\let\useblackboard=\iffalse
%
% You might also need to remove this line.
\newfam\black

\input harvmac
%\input epsf.tex

%%BLACKBOARD FONT STUFF
\useblackboard
\message{If you do not have msbm (blackboard bold) fonts,}
\message{change the option at the top of the tex file.}
%Why is this magstep1? It makes the bb font bigger than rm --MBS
%You're asking me?  I stole this source code from your mom -- AEL.
%\font\blackboard=msbm10 scaled \magstep1
\font\blackboard=msbm10
\font\blackboards=msbm7
\font\blackboardss=msbm5
\textfont\black=\blackboard
\scriptfont\black=\blackboards
\scriptscriptfont\black=\blackboardss
\def\Bbb#1{{\fam\black\relax#1}}
\else
\def\Bbb{\bf}
\fi
%Macros for boxes from cordes moore ramgoolam
\def\boxit#1{\vbox{\hrule\hbox{\vrule\kern8pt
\vbox{\hbox{\kern8pt}\hbox{\vbox{#1}}\hbox{\kern8pt}}
\kern8pt\vrule}\hrule}}
\def\mathboxit#1{\vbox{\hrule\hbox{\vrule\kern8pt\vbox{\kern8pt
\hbox{$\displaystyle #1$}\kern8pt}\kern8pt\vrule}\hrule}}

%% for sub sub sections("...poor devil of a sub-sub...")
\def\subsubsec#1{\ifnum\lastpenalty>9000\else\bigbreak\fi
\noindent{\it{#1}}\par\nobreak\medskip\nobreak}
\def\yboxit#1#2{\vbotx{\hrule height #1 \hbox{\vrule width #1
\vbox{#2}\vrule width #1 }\hrule height #1 }}
\def\fillbox#1{\hbox to #1{\vbox to #1{\vfil}\hfil}}
\def\ybox{{\lower 1.3pt \yboxit{0.4pt}{\fillbox{8pt}}\hskip-0.2pt}}

\def\QR{\Bbb{R}}

\def\QZ{\Bbb{Z}}

\def\eps{\epsilon}

\def\bthet{\bar \theta }

%''We will all char together when we char
%And let there be no moaning at the bar''
%Tom Lehrer

\def\jb{{\bar\jmath}}
\def\bz{{\bar z}}

\def\bhthet{\hat{\bthet}}

\def\CN{{\cal N}}
\def\CW{{\cal W}}
\def\N{{\cal N}}

\def\CH{{\cal H}}
\def\CD{{\cal D}}

\

%  draw box of size #1pt and line thickness #2pt
\def\drawbox#1#2{\hrule height#2pt 
        \hbox{\vrule width#2pt height#1pt \kern#1pt \vrule width#2pt}
              \hrule height#2pt}
% Young tableaux

\def\Asym#1#2{\vcenter{\vbox{\drawbox{#1}{#2}
              \kern-#2pt       % line up boxes
              \drawbox{#1}{#2}}}}

%%More math defs
\def\p{\partial}

\def\frac#1#2{{#1 \over #2}}
\def\Del{\nabla}

%%bars
\def\jb{{\bar\jmath}}

%%misc
\def\re{{\rm Re\ }}

\def\cf{{\rm cf.}}  % It's cf. not c.f. since `conferre' (latin) is one word.
 % These abbrevs. are so common that they're not italicized. 
\def\Spin{{\rm Spin}}

%% Optional changes in style
\def\NSNS{NS-NS}  %NSNS or NS-NS?
\def\thypermultiplet{hypermultiplet} % was ``tensor hypermultiplet''
\def\thypermultiplets{hypermultiplets} % was ``tensor hypermultiplets''
\def\Thypermultiplets{Hypermultiplets} % was ``Tensor hypermultiplets''

%%Refs
%Nongeometry for dummies
\lref\HullIN{
  C.~M.~Hull,
  ``A geometry for non-geometric string backgrounds,''
  JHEP {\bf 0510}, 065 (2005)
  [arXiv:hep-th/0406102].
  %%CITATION = HEP-TH 0406102;%%
}
\lref\FlournoyVN{
  A.~Flournoy, B.~Wecht and B.~Williams,
  ``Constructing nongeometric vacua in string theory,''
  Nucl.\ Phys.\ B {\bf 706}, 127 (2005)
  [arXiv:hep-th/0404217].
  %%CITATION = HEP-TH 0404217;%%
}
%\HellermanAX
\lref\HellermanAX{
  S.~Hellerman, J.~McGreevy and B.~Williams,
  ``Geometric constructions of nongeometric string theories,''
  JHEP {\bf 0401}, 024 (2004)
  [arXiv:hep-th/0208174].
  %%CITATION = HEP-TH 0208174;%%
}
%\SheltonCF
\lref\SheltonCF{
  J.~Shelton, W.~Taylor and B.~Wecht,
  ``Nongeometric flux compactifications,''
  JHEP {\bf 0510}, 085 (2005)
  [arXiv:hep-th/0508133].
  %%CITATION = HEP-TH 0508133;%%
}

%%Generalized Calabi-Yauism
%\HitchinUT
\lref\HitchinUT{
  N.~Hitchin,
  ``Generalized Calabi-Yau manifolds,''
  Quart.\ J.\ Math.\ Oxford Ser.\  {\bf 54}, 281 (2003)
  [arXiv:math.dg/0209099].
  %%CITATION = MATH-DG 0209099;%%
}
%\GualtieriThesis
\lref\GualtieriThesis{
 M.~Gualtieri,
 ``Generalized complex geometry",
 arXiv:math.DG/0401221.
}
%\GrangeNM
\lref\GrangeNM{
  P.~Grange and R.~Minasian,
  ``Tachyon condensation and D-branes in generalized geometries,''
  arXiv:hep-th/0512185.
  %%CITATION = HEP-TH 0512185;%%
}

%%all is in flux
%\GukovYA
\lref\GukovYA{
  S.~Gukov, C.~Vafa and E.~Witten,
  ``CFT's from Calabi-Yau four-folds,''
  Nucl.\ Phys.\ B {\bf 584}, 69 (2000)
  [Erratum-ibid.\ B {\bf 608}, 477 (2001)]
  [arXiv:hep-th/9906070].
  %%CITATION = HEP-TH 9906070;%%
}
%\TaylorII
\lref\TaylorII{
  T.~R.~Taylor and C.~Vafa,
  ``RR flux on Calabi-Yau and partial supersymmetry breaking,''
  Phys.\ Lett.\ B {\bf 474}, 130 (2000)
  [arXiv:hep-th/9912152].
  %%Cited 237 times in SPIRES-HEP
}
%\VafaWI
\lref\VafaWI{
  C.~Vafa,
  ``Superstrings and topological strings at large N,''
  J.\ Math.\ Phys.\  {\bf 42}, 2798 (2001)
  [arXiv:hep-th/0008142].
  %%CITATION = HEP-TH 0008142;%%
}
%\LawrenceZK
\lref\LawrenceZK{
  A.~Lawrence and J.~McGreevy,
  ``Local string models of soft supersymmetry breaking,''
  JHEP {\bf 0406}, 007 (2004)
  [arXiv:hep-th/0401034].
  %%CITATION = HEP-TH 0401034;%%
}
%\LawrenceKJ
\lref\LawrenceKJ{
  A.~Lawrence and J.~McGreevy,
  ``Remarks on branes, fluxes, and soft SUSY breaking,''
  arXiv:hep-th/0401233.
  %%CITATION = HEP-TH 0401233;%%
}

%%Heterotic 
%\BeckerYV
\lref\BeckerYV{
  K.~Becker, M.~Becker, K.~Dasgupta and P.~S.~Green,
  ``Compactifications of heterotic theory on non-Kaehler complex manifolds.
  I,''
  JHEP {\bf 0304}, 007 (2003)
  [arXiv:hep-th/0301161].
  %%CITATION = HEP-TH 0301161;%%
  %%Cited 74 times in SPIRES-HEP
}

%%Mirror symmetry
%\StromingerIT
\lref\StromingerIT{
  A.~Strominger, S.~T.~Yau and E.~Zaslow,
  ``Mirror symmetry is T-duality,''
  Nucl.\ Phys.\ B {\bf 479}, 243 (1996)
  [arXiv:hep-th/9606040].
  %%CITATION = HEP-TH 9606040;%%
}

%%Fun with fluxes
%\KachruSK
\lref\KachruSK{
  S.~Kachru, M.~B.~Schulz, P.~K.~Tripathy and S.~P.~Trivedi,
  ``New supersymmetric string compactifications,''
  JHEP {\bf 0303}, 061 (2003)
  [arXiv:hep-th/0211182].
  %%CITATION = HEP-TH 0211182;%%
}
\lref\KachruHE{
  S.~Kachru, M.~B.~Schulz and S.~Trivedi,
  ``Moduli stabilization from fluxes in a simple IIB orientifold,''
  JHEP {\bf 0310}, 007 (2003)
  [arXiv:hep-th/0201028].
  %%CITATION = HEP-TH 0201028;%%
}
\lref\DeWolfeUU{
  O.~DeWolfe, A.~Giryavets, S.~Kachru and W.~Taylor,
  ``Type IIA moduli stabilization,''
  JHEP {\bf 0507}, 066 (2005)
  [arXiv:hep-th/0505160].
  %%CITATION = HEP-TH 0505160;%%
}
\lref\DabholkarSY{
  A.~Dabholkar and C.~Hull,
  ``Duality twists, orbifolds, and fluxes,''
  JHEP {\bf 0309}, 054 (2003)
  [arXiv:hep-th/0210209].
  %%CITATION = HEP-TH 0210209;%%
}
\lref\DabholkarVE{
  A.~Dabholkar and C.~Hull,
  ``Generalised T-duality and non-geometric backgrounds,''
  arXiv:hep-th/0512005.
  %%CITATION = HEP-TH 0512005;%%
}
\lref\HullHK{
  C.~M.~Hull and R.~A.~Reid-Edwards,
  ``Flux compactifications of string theory on twisted tori,''
  arXiv:hep-th/0503114.
  %%CITATION = HEP-TH 0503114;%%
}
\lref\GrayEA{
  J.~Gray and E.~Hackett-Jones,
  ``On T-folds, G-structures and supersymmetry,''
  arXiv:hep-th/0506092.
  %%CITATION = HEP-TH 0506092;%%
}
%\GranaSN
\lref\GranaSN{
M.~Gra\~na, R.~Minasian, M.~Petrini and A.~Tomasiello,
``Generalized structures of N = 1 vacua,''
JHEP {\bf 0511}, 020 (2005)
[arXiv:hep-th/0505212].
%%CITATION = HEP-TH 0505212;%%
}
%\GranaSV
\lref\GranaSV{
M.~Gra\~na, R.~Minasian, M.~Petrini and A.~Tomasiello,
``Type II strings and generalized Calabi-Yau manifolds,''
Comptes Rendus Physique {\bf 5}, 979 (2004)
[arXiv:hep-th/0409176].
%%CITATION = HEP-TH 0409176;%%
}
%\GranaBG
\lref\GranaBG{
M.~Gra\~na, R.~Minasian, M.~Petrini and A.~Tomasiello,
``Supersymmetric backgrounds from generalized Calabi-Yau manifolds,''
JHEP {\bf 0408}, 046 (2004)
[arXiv:hep-th/0406137].
%%CITATION = HEP-TH 0406137;%%
}
%\KashaniPoorSI
\lref\KashaniPoorSI{
  A.~K.~Kashani-Poor and R.~Minasian,
  ``Towards reduction of type II theories on SU(3) structure manifolds,''
  JHEP {\bf 0703}, 109 (2007)
  [arXiv:hep-th/0611106].
  %%CITATION = JHEPA,0703,109;%%
}
\lref\GurrieriIW{
  S.~Gurrieri and A.~Micu,
  ``Type IIB theory on half-flat manifolds,''
  Class.\ Quant.\ Grav.\  {\bf 20}, 2181 (2003)
  [arXiv:hep-th/0212278].
  %%CITATION = HEP-TH 0212278;%%
}
\lref\GurrieriDT{
  S.~Gurrieri, A.~Lukas and A.~Micu,
  ``Heterotic on half-flat,''
  Phys.\ Rev.\ D {\bf 70}, 126009 (2004)
  [arXiv:hep-th/0408121].
  %%CITATION = HEP-TH 0408121;%%
}
\lref\FidanzaZI{
  S.~Fidanza, R.~Minasian and A.~Tomasiello,
  ``Mirror symmetric SU(3)-structure manifolds with NS fluxes,''
  Commun.\ Math.\ Phys.\  {\bf 254}, 401 (2005)
  [arXiv:hep-th/0311122].
  %%CITATION = HEP-TH 0311122;%%
  %%Cited 50 times in SPIRES-HEP
}
\lref\BeckerII{
  K.~Becker, M.~Becker, K.~Dasgupta and R.~Tatar,
  ``Geometric transitions, non-Kaehler geometries and string vacua,''
  Int.\ J.\ Mod.\ Phys.\ A {\bf 20}, 3442 (2005)
  [arXiv:hep-th/0411039].
  %%CITATION = HEP-TH 0411039;%%
  %%Cited 1 time in SPIRES-HEP
}
\lref\AlexanderEQ{
  S.~Alexander, K.~Becker, M.~Becker, K.~Dasgupta, A.~Knauf and R.~Tatar,
  ``In the realm of the geometric transitions,''
  Nucl.\ Phys.\ B {\bf 704}, 231 (2005)
  [arXiv:hep-th/0408192].
  %%CITATION = HEP-TH 0408192;%%
  %%Cited 9 times in SPIRES-HEP
}
%\TomasielloBP
\lref\TomasielloBP{
A.~Tomasiello,
``Topological mirror symmetry with fluxes,''
JHEP {\bf 0506}, 067 (2005)
[arXiv:hep-th/0502148].
%%CITATION = HEP-TH 0502148;%%
}

\lref\GurrieriWZ{
  S.~Gurrieri, J.~Louis, A.~Micu and D.~Waldram,
  ``Mirror symmetry in generalized Calabi-Yau compactifications,''
  Nucl.\ Phys.\ B {\bf 654}, 61 (2003)
  [arXiv:hep-th/0211102].
  %%CITATION = HEP-TH 0211102;%%
}

%\GranaNY
\lref\GranaNY{
  M.~Gra\~na, J.~Louis and D.~Waldram,
  ``Hitchin functionals in N = 2 supergravity,''
  JHEP {\bf 0601}, 008 (2006)
  [arXiv:hep-th/0505264].
  %%CITATION = JHEPA,0601,008;%%
}

%Worldsheet instantons
%\DineZY
\lref\DineZY{
  M.~Dine, N.~Seiberg, X.~G.~Wen and E.~Witten,
  ``Nonperturbative effects on the string worldsheet,''
  Nucl.\ Phys.\ B {\bf 278}, 769 (1986).
  %%CITATION = NUPHA,B278,769;%%
}
%\DineBQ
\lref\DineBQ{
  M.~Dine, N.~Seiberg, X.~G.~Wen and E.~Witten,
  ``Nonperturbative effects on the string worldsheet. 2,''
  Nucl.\ Phys.\ B {\bf 289}, 319 (1987).
  %%CITATION = NUPHA,B289,319;%%
}
%\KachruIH
\lref\KachruIH{
  S.~Kachru, S.~Katz, A.~E.~Lawrence and J.~McGreevy,
  ``Open string instantons and superpotentials,''
  Phys.\ Rev.\ D {\bf 62}, 026001 (2000)
  [arXiv:hep-th/9912151].
  %%CITATION = HEP-TH 9912151;%%
}
%\ChamseddineMC
\lref\ChamseddineMC{
  A.~H.~Chamseddine and M.~S.~Volkov,
  ``Non-Abelian solitons in N = 4 gauged supergravity and leading order  string
  theory,''
  Phys.\ Rev.\  D {\bf 57}, 6242 (1998)
  [arXiv:hep-th/9711181].
  %%CITATION = PHRVA,D57,6242;%%
}
%\MaldacenaYY
\lref\MaldacenaYY{
  J.~M.~Maldacena and C.~Nunez,
  ``Towards the large N limit of pure N = 1 super Yang Mills,''
  Phys.\ Rev.\ Lett.\  {\bf 86}, 588 (2001)
  [arXiv:hep-th/0008001].
  %%CITATION = HEP-TH 0008001;%%
}

%\BanksCY
\lref\BanksCY{
  T.~Banks, L.~J.~Dixon, D.~Friedan and E.~J.~Martinec,
   ``Phenomenology and conformal field theory or can string theory
   predict the weak mixing angle?,''
  Nucl.\ Phys.\ B {\bf 299}, 613 (1988).
  %%CITATION = NUPHA,B299,613;%%
}
%\BanksYZ
\lref\BanksYZ{
  T.~Banks and L.~J.~Dixon,
  ``Constraints on string vacua with spacetime supersymmetry,''
  Nucl.\ Phys.\ B {\bf 307}, 93 (1988).
  %%CITATION = NUPHA,B307,93;%%
}

%D-terms
%%References
%\LawrenceSM
\lref\LawrenceSM{
  A.~Lawrence and J.~McGreevy,
   ``D-terms and D-strings in open string models,''
  JHEP {\bf 0410}, 056 (2004)
  [arXiv:hep-th/0409284].
  %%CITATION = HEP-TH 0409284;%%
}
%\BerkovitsCB
\lref\BerkovitsCB{
  N.~Berkovits and W.~Siegel,
   ``Superspace Effective Actions for 4D Compactifications of
   Heterotic and Type II Superstrings,''
  Nucl.\ Phys.\ B {\bf 462}, 213 (1996)
  [arXiv:hep-th/9510106].
  %%CITATION = HEP-TH 9510106;%%
}
\lref\ChiossiSal{
  S. Choissi and S. Salamon,  ``The Intrinsic Torsion of SU(3)
  and $G_2$ structures," 
  [arxiv:math.DG/0202282]}
%\GrimmXP
\lref\GrimmXP{
  R.~Grimm, M.~Sohnius and J.~Wess,
  ``Extended Supersymmetry And Gauge Theories,''
  Nucl.\ Phys.\ B {\bf 133}, 275 (1978).
  %%CITATION = NUPHA,B133,275;%%
}
%\deWitPQ
\lref\deWitPQ{
  B.~de Wit and J.~W.~van Holten,
  ``Multiplets Of Linearized SO(2) Supergravity,''
  Nucl.\ Phys.\ B {\bf 155}, 530 (1979).
  %%CITATION = NUPHA,B155,530;%%
}
%\deRooMM
\lref\deRooMM{
  M.~de Roo, J.~W.~van Holten, B.~de Wit and A.~Van Proeyen,
  ``Chiral Superfields In N=2 Supergravity,''
  Nucl.\ Phys.\ B {\bf 173}, 175 (1980).
  %%CITATION = NUPHA,B173,175;%%
}
%\GirardelloWZ
\lref\GirardelloWZ{
  L.~Girardello and M.~T.~Grisaru,
  ``Soft Breaking Of Supersymmetry,''
  Nucl.\ Phys.\ B {\bf 194}, 65 (1982).
  %%CITATION = NUPHA,B194,65;%%
}
%\AtickGY
\lref\AtickGY{
  J.~J.~Atick, L.~J.~Dixon and A.~Sen,
   ``String Calculation Of Fayet-Iliopoulos D Terms In Arbitrary
   Supersymmetric Compactifications,''
  Nucl.\ Phys.\ B {\bf 292}, 109 (1987).
  %%CITATION = NUPHA,B292,109;%%
}
%\DineGJ
\lref\DineGJ{
  M.~Dine, I.~Ichinose and N.~Seiberg,
  ``F Terms And D Terms In String Theory,''
  Nucl.\ Phys.\ B {\bf 293}, 253 (1987).
  %%CITATION = NUPHA,B293,253;%%
}

%\LawrenceMA
\lref\LawrenceMA{
  A.~Lawrence, M.~B.~Schulz and B.~Wecht,
  ``D-branes in nongeometric backgrounds,''
  JHEP {\bf 0607}, 038 (2006)
  [arXiv:hep-th/0602025].
  %%CITATION = JHEPA,0607,038;%%
}
%\GatesNK
\lref\GatesNK{
  S.~J.~Gates, C.~M.~Hull and M.~Rocek,
  ``Twisted Multiplets And New Supersymmetric Nonlinear Sigma Models,''
  Nucl.\ Phys.\  B {\bf 248}, 157 (1984).
  %%CITATION = NUPHA,B248,157;%%
}
%\DouglasGI
\lref\DouglasGI{
  M.~R.~Douglas,
  ``D-branes, categories and N = 1 supersymmetry,''
  J.\ Math.\ Phys.\  {\bf 42}, 2818 (2001)
  [arXiv:hep-th/0011017].
  %%CITATION = JMAPA,42,2818;%%
}
%\HullZY
\lref\HullZY{
  C.~M.~Hull,
  ``Sigma Model Beta Functions And String Compactifications,''
  Nucl.\ Phys.\  B {\bf 267}, 266 (1986).
  %%CITATION = NUPHA,B267,266;%%
}
%\AlvarezGaumeHN
\lref\AlvarezGaumeHN{
  L.~Alvarez-Gaum\'e, D.~Z.~Freedman and S.~Mukhi,
  ``The Background Field Method And The Ultraviolet Structure Of The
  Supersymmetric Nonlinear Sigma Model,''
  Annals Phys.\  {\bf 134}, 85 (1981).
  %%CITATION = APNYA,134,85;%%
}
%\LawrenceBK
\lref\LawrenceBK{
  A.~Lawrence and A.~Sever,
  ``Scattering of twist fields from D-branes and orientifolds,''
  arXiv:0706.3199 [hep-th].
  %%CITATION = ARXIV:0706.3199;%%
}
%\GranaHR
\lref\GranaHR{
  M.~Gra\~na, J.~Louis and D.~Waldram,
  ``SU(3) x SU(3) compactification and mirror duals of magnetic fluxes,''
  JHEP {\bf 0704}, 101 (2007)
  [arXiv:hep-th/0612237].
  %%CITATION = JHEPA,0704,101;%%
}

%\GranaKF
\lref\GranaKF{
  M.~Gra\~na, R.~Minasian, M.~Petrini and A.~Tomasiello,
  ``A scan for new N=1 vacua on twisted tori,''
  JHEP {\bf 0705}, 031 (2007)
  [arXiv:hep-th/0609124].
  %%CITATION = JHEPA,0705,031;%%
}

%\SheltonFD
\lref\SheltonFD{
  J.~Shelton, W.~Taylor and B.~Wecht,
  ``Generalized flux vacua,''
  JHEP {\bf 0702}, 095 (2007)
  [arXiv:hep-th/0607015].
  %%CITATION = JHEPA,0702,095;%%
}
%\WechtWU
\lref\WechtWU{
  B.~Wecht,
  ``Lectures on Nongeometric Flux Compactifications,''
  arXiv:0708.3984 [hep-th].
  %%CITATION = ARXIV:0708.3984;%%
}
%\MichelsonPN
\lref\MichelsonPN{
  J.~Michelson,
  ``Compactifications of type IIB strings to four dimensions with  non-trivial
  classical potential,''
  Nucl.\ Phys.\  B {\bf 495}, 127 (1997)
  [arXiv:hep-th/9610151].
  %%CITATION = NUPHA,B495,127;%%
}
%\KachruEM
\lref\KachruEM{
  S.~Kachru, J.~McGreevy and P.~Svrcek,
  ``Bounds on masses of bulk fields in string compactifications,''
  JHEP {\bf 0604}, 023 (2006)
  [arXiv:hep-th/0601111].
  %%CITATION = JHEPA,0604,023;%%
}
%\DeWolfeNN
\lref\DeWolfeNN{
  O.~DeWolfe and S.~B.~Giddings,
  ``Scales and hierarchies in warped compactifications and brane worlds,''
  Phys.\ Rev.\  D {\bf 67}, 066008 (2003)
  [arXiv:hep-th/0208123].
  %%CITATION = PHRVA,D67,066008;%%
}
%\BanksDH
\lref\BanksDH{
  T.~Banks,
  ``Remarks on M theoretic cosmology,''
  arXiv:hep-th/9906126.
  %%CITATION = HEP-TH/9906126;%%
}
%\BanksAY
\lref\BanksAY{
  T.~Banks,
  ``M-theory and cosmology,''
  arXiv:hep-th/9911067.
  %%CITATION = HEP-TH/9911067;%%
}
%\HoravaQA
\lref\HoravaQA{
  P.~Horava and E.~Witten,
  ``Heterotic and type I string dynamics from eleven dimensions,''
  Nucl.\ Phys.\  B {\bf 460}, 506 (1996)
  [arXiv:hep-th/9510209].
  %%CITATION = NUPHA,B460,506;%%
}
%\HoravaMA
\lref\HoravaMA{
  P.~Horava and E.~Witten,
  ``Eleven-Dimensional Supergravity on a Manifold with Boundary,''
  Nucl.\ Phys.\  B {\bf 475}, 94 (1996)
  [arXiv:hep-th/9603142].
  %%CITATION = NUPHA,B475,94;%%
}
%\PolchinskiRQ
\lref\PolchinskiRQ{
  J.~Polchinski,
  ``String theory. Vol. 1: An introduction to the bosonic string,''
%\href{http://www.slac.stanford.edu/spires/find/hep/www?irn=4634799}{SPIRES entry}
{\it  Cambridge, UK: Univ. Pr. (1998) 402 p}
}
%\PolchinskiRR
\lref\PolchinskiRR{
  J.~Polchinski,
  ``String theory. Vol. 2: Superstring theory and beyond,''
%\href{http://www.slac.stanford.edu/spires/find/hep/www?irn=4634802}{SPIRES entry}
{\it  Cambridge, UK: Univ. Pr. (1998) 531 p}
}
%\GranaJC
\lref\GranaJC{
  M.~Gra\~na,
  ``Flux compactifications in string theory: A comprehensive review,''
  Phys.\ Rept.\  {\bf 423}, 91 (2006)
  [arXiv:hep-th/0509003].
  %%CITATION = PRPLC,423,91;%%
}
%\MorrisonBT
\lref\MorrisonBT{
  D.~R.~Morrison,
  ``Geometric aspects of mirror symmetry,''
  arXiv:math/0007090.
  %%CITATION = MATH/0007090;%%
}
%\MicuRD
\lref\MicuRD{
  A.~Micu, E.~Palti and G.~Tasinato,
  ``Towards Minkowski vacua in type II string compactifications,''
  JHEP {\bf 0703}, 104 (2007)
  [arXiv:hep-th/0701173].
  %%CITATION = JHEPA,0703,104;%%
}
%\BenmachicheDF
\lref\BenmachicheDF{
  I.~Benmachiche and T.~W.~Grimm,
  ``Generalized N = 1 orientifold compactifications and the Hitchin
  functionals,''
  Nucl.\ Phys.\  B {\bf 748}, 200 (2006)
  [arXiv:hep-th/0602241].
  %%CITATION = NUPHA,B748,200;%%
}
%\AldazabalUP
\lref\AldazabalUP{
  G.~Aldazabal, P.~G.~Camara, A.~Font and L.~E.~Ibanez,
  ``More dual fluxes and moduli fixing,''
  JHEP {\bf 0605}, 070 (2006)
  [arXiv:hep-th/0602089].
  %%CITATION = JHEPA,0605,070;%%
}
%\KachruNS
\lref\KachruNS{
  S.~Kachru, X.~Liu, M.~B.~Schulz and S.~P.~Trivedi,
  ``Supersymmetry changing bubbles in string theory,''
  JHEP {\bf 0305}, 014 (2003)
  [arXiv:hep-th/0205108].
  %%CITATION = JHEPA,0305,014;%%
}
%\LindstromKS
\lref\LindstromKS{
  U.~Lindstrom and M.~Rocek,
  ``New hyperkahler metrics and new supermultiplets,''
  Commun.\ Math.\ Phys.\  {\bf 115}, 21 (1988).
  %%CITATION = CMPHA,115,21;%%
}
%\GranaNQ
\lref\GranaNQ{
  M.~Gra\~na,
  ``MSSM parameters from supergravity backgrounds,''
  Phys.\ Rev.\  D {\bf 67}, 066006 (2003)
  [arXiv:hep-th/0209200].
  %%CITATION = PHRVA,D67,066006;%%
}
%\CamaraKU
\lref\CamaraKU{
  P.~G.~Camara, L.~E.~Ibanez and A.~M.~Uranga,
  ``Flux-induced SUSY-breaking soft terms,''
  Nucl.\ Phys.\  B {\bf 689}, 195 (2004)
  [arXiv:hep-th/0311241].
  %%CITATION = NUPHA,B689,195;%%
}
%\CamaraCZ
\lref\CamaraCZ{
  P.~G.~Camara and M.~Gra\~na,
  ``No-scale supersymmetry breaking vacua and soft terms with torsion,''
  arXiv:0710.4577 [hep-th].
  %%CITATION = ARXIV:0710.4577;%%
}
\lref\DenefMM{
  F.~Denef, M.~R.~Douglas, B.~Florea, A.~Grassi and S.~Kachru,
  ``Fixing all moduli in a simple F-theory compactification,''
  Adv.\ Theor.\ Math.\ Phys.\  {\bf 9}, 861 (2005)
  [arXiv:hep-th/0503124].
  %%CITATION = 00203,9,861;%%
}

%Asymmetric Orbifolds and other nongeo
%\FlournoyXE
\lref\FlournoyXE{
  A.~Flournoy and B.~Williams,
  ``Nongeometry, duality twists, and the worldsheet,''
  arXiv:hep-th/0511126.
  %%CITATION = HEP-TH 0511126;%%
}
%\SilversteinID
\lref\SilversteinID{
  E.~Silverstein,
  ``TASI / PiTP / ISS lectures on moduli and microphysics,''
  arXiv:hep-th/0405068.
  %%CITATION = HEP-TH/0405068;%%
}
%\MaloneyRR
\lref\MaloneyRR{
  A.~Maloney, E.~Silverstein and A.~Strominger,
  ``De Sitter space in noncritical string theory,''
  arXiv:hep-th/0205316.
  %%CITATION = HEP-TH/0205316;%%
}
%\SilversteinXN
\lref\SilversteinXN{
  E.~Silverstein,
  ``(A)dS backgrounds from asymmetric orientifolds,''
  arXiv:hep-th/0106209.
  %%CITATION = HEP-TH/0106209;%%
}
%\HellermanTX
\lref\HellermanTX{
  S.~Hellerman and J.~Walcher,
  ``Worldsheet CFTs for flat monodrofolds,''
  arXiv:hep-th/0604191.
  %%CITATION = HEP-TH/0604191;%%
}
%\BeckerKS
\lref\BeckerKS{
  K.~Becker, M.~Becker, C.~Vafa and J.~Walcher,
  ``Moduli stabilization in non-geometric backgrounds,''
  Nucl.\ Phys.\  B {\bf 770}, 1 (2007)
  [arXiv:hep-th/0611001].
  %%CITATION = NUPHA,B770,1;%%
}
\lref\amiretal{
J.~de Boer, A.~Kashani-Poor, and S.~El-Showk, to appear.}

\Title{\vbox{\baselineskip12pt
\hbox{NSF-KITP-06-97}
\hbox{MIT-CTP-3798}
\hbox{BRX TH-578}
\hbox{UPR-1176-T}}}
{\vbox{\centerline{Torsion and Supersymmetry Breaking}}}
\vskip -2em
\centerline{Albion Lawrence${}^{1}$, 
Tobias Sander${}^1$,}
\centerline{ Michael B. Schulz${}^{2,3}$, and Brian
Wecht${}^{4,5}$} 
\medskip
\centerline{${}^1$ {\it Theory Group, Martin Fisher School of
Physics, Brandeis University,}} 
\centerline{{\it MS 057, PO Box 549110, Waltham, MA 02454, USA}}
\centerline{${}^2$ {\it Department of Physics, Bryn Mawr College, Bryn Mawr, PA 19010, USA}}
\centerline{${}^3$ {\it Department of Physics and Astronomy,
University of Pennsylvania, Philadelphia, PA 19104, USA}}
\centerline{${}^4$ {\it Center for Theoretical Physics, MIT, Cambridge, MA 02139, USA}}
\centerline{${}^5$ {\it School of Natural Sciences, Institute for Advanced Study, Princeton,
NJ 08450, USA}}

\bigskip
\noindent
We identify the auxiliary fields in the hypermultiplets of type IIB
string theory compactified on a Calabi-Yau manifold, using a
combination of worldsheet and supergravity techniques.  
The SUSY-breaking squark and gaugino masses
in type IIB models depend on these auxiliary fields,
which parametrize deformations away from a pure
Calabi-Yau compactification to one with \NSNS\ 3-form flux and
$SU(3)\times SU(3)$ structure.  Worldsheet arguments show that such
compactifications are generically globally nongeometric. Our results,
combined with earlier results for type IIA compactifications, imply
that these deformations are the mirrors of \NSNS\ 3-form flux, in
accord with work from the supergravity point of view. Using the
worldsheet current algebra, we explain why mirror symmetry may
continue to hold in the presence of fluxes breaking the symmetries
(e.g., (2,2) SUSY) on which mirror symmetry is typically taken to
depend.  Finally, we give evidence that nonperturbative worldsheet
effects (such as worldsheet instantons) provide important corrections
to the supergravity picture in the presence of auxiliary fields for
K\"ahler moduli.

\medskip
%\draftmode
\Date{\number\day\ November 2007}

\listtoc
\writetoc

\newsec{Introduction}

In this article, we compute the auxiliary fields of $\CN=2$
hypermultiplets in type IIB Calabi-Yau compactifications to four
dimensions.  In type IIB, these multiplets contain the K\"ahler moduli
and the dilaton-axion, as well as the RR axions.  This work is a
continuation of the program begun in \refs{\LawrenceZK,\LawrenceKJ},
which focused on the vector multiplets.

Our work has several motivations.  The first, stemming from particle
physics model building considerations, is that expectation values for
these auxiliary fields generate explicit SUSY-breaking terms in the
low energy four dimensional theory \refs{\GirardelloWZ}. This has
proven useful for understanding $\CN=1$ flux compactifications.  For
example, it was shown in \refs{\LawrenceZK,\LawrenceKJ,\VafaWI}\ that
the flux-induced superpotential $W = \int G_3 \wedge \Omega$ derived
in \refs{\GukovYA,\TaylorII,\MichelsonPN}\ can be computed as a term
explicitly breaking $\CN=2$ to $\CN=1$ supersymmetry, proportional to
the expectation values of auxiliary fields in vector multiplets.

The same computations are also useful in studying $\CN = 1$ to $\CN =
0$ supersymmetry breaking at lower energies in models with D-branes
and/or fluxes.  If $\CN=2$ supersymmetry is broken at a higher scale
than $\CN=1$, as happens in many flux compactifications, and
$N=1$ supersymmetry is broken at a still lower scale, one can
separate the auxiliary fields into those (call them $F_{\rm high}$)
whose expectation values break $\CN=2$ to $\CN=1$, and those (call
them $F_{\rm low}$) whose expectation values break $\CN=1$
to \hbox{$\CN=0$}.  After performing the necessary orientifold
projections compatible with the $\CN=1$ supersymmetry, our results
should allow the closed string fields to be written as $\CN=1$
superfields with auxiliary fields of type $F_{\rm low}$.  The
auxiliary fields in these supermultiplets then parametrize supersymmetry
breaking in the low-energy $\CN=1$ effective Lagrangian.  In
particular, the SUSY-breaking squark and gaugino
masses will depend on the auxiliary fields we compute here 
\refs{\LawrenceZK,\LawrenceKJ}. In the conclusions we will
discuss some recent work on building SUSY-breaking string models,
for which the results here and in \refs{\LawrenceZK,\LawrenceKJ}\ 
have some relevance.

A second motivation for our work arises from the desire to extend
the powerful results of mirror symmetry to compactifications with $\CN < 2$ 
spacetime supersymmetry.  For type II compactifications these involve
NS-NS fluxes, and finding the mirrors of compactifications with such fluxes is a long-standing
problem. However, in the cases that we can understand a compactification with $\CN < 2$
supersymmetry as a deformation of a $\CN=2$ compactification by 
expectation values for auxiliary fields, we can make progress
by understanding the action of the mirror map on these auxiliary fields.

More precisely, the auxiliary fields for hypermultiplets in type IIA (whose scalar components
include the complex structure moduli) have been identified in
\refs{\LawrenceZK,\LawrenceKJ} as a combination of \NSNS\ 3-form
flux and a subset of the $SU(3)$ intrinsic torsion classes.\foot{In
the notation of \refs{\ChiossiSal}, the relevant torsion classes are
$W_3$ and $W_4$, defined in Sec.~2.4 below.}  These torsion classes
parametrize deformations of the compactification away from a pure
special holonomy compactification. The mirrors of these fluxes and
torsion classes should be the auxiliary fields for the type IIB
hypermultiplets, which include the K\"ahler moduli.

Supergravity arguments already suggest an answer.
The mirrors of type IIA compactifications with purely
electric\foot{Here, purely electric NS-NS flux means flux through the
$A$ cycles but not $B$ cycles, in a symplectic basis of $H_3$.} \NSNS\
flux in $H^{(2,1)}(X)\oplus H^{(1,2)}(X)$ and intrinsic torsion of
type $W_{3,4}$, have been identified with ``half-flat" manifolds
\refs{\GurrieriWZ\GranaNY\GranaSN\GranaSV\GranaBG
\GurrieriIW\FidanzaZI\TomasielloBP-\GurrieriDT}.  For more general NS-NS flux, the
mirrors have been identified with compactifications of $SU(3)\times
SU(3)$ structure \refs{\GranaNY,\GranaSN,\GranaHR}.\foot{Here,
$SU(3)\times SU(3)$ refers to the structure group of an extension of
the bundle $T\oplus T^*$, or equivalently, to distinct left and right
moving $SU(3)$ structures of the usual frame bundle.} On the other
hand, considerations of the effective four-dimensional
superpotential \refs{\SheltonCF\SheltonFD \WechtWU\AldazabalUP\BenmachicheDF-\MicuRD}\ 
and of the action of \hbox{T-duality} transformations (such as mirror symmetry) on \NSNS\ flux
\refs{\KachruSK\DabholkarSY\DabholkarVE\HullHK-\GrayEA}\ indicate
that these mirrors should be generically nongeometric in the sense
discussed in \refs{\HellermanAX} (other related approaches to nongeometric backgrounds include asymmetric orbifolds \refs{\FlournoyXE\HellermanTX\SilversteinID\MaloneyRR-\SilversteinXN} and Landau-Ginzburg models \BeckerKS). One goal of the present work is to
make this claim more precise, and to relate auxiliary fields in IIB
hypermultiplets to (nongeometric) intrinsic torsions in 
$SU(3) \times SU(3)$ structure compactifications. 

Our basic approach uses an $\CN=2$ superspace formalism that is
natural from the worldsheet point of view.  Our computations confirm
{\it both}\ lessons of the previous paragraph. Auxiliary fields for
K\"ahler moduli correspond locally on the target space to intrinsic
torsion classes for background with $SU(3)\times SU(3)$ structure.
However, when the two auxiliary fields in a given multiplet are
dialed independently of each other, the string background is generically
nongeometric.\foot{Note that Refs.~\refs{\SheltonCF,\SheltonFD,\WechtWU} also
consider backgrounds with ``$R$-flux,'' which are not even locally
geometric.  We suspect that we are missing the fluxes
because we are considering deformations of a geometric background.}
This need for nongeometric structures becomes clear from the
worldsheet, as we will discuss. Furthermore, one can understand why
mirror symmetry may still be valid from the worldsheet point of view:
it corresponds to reversing the sign of a $U(1)_R$ current, which
exists even though it is no longer conserved.

Note that while the torsion classes are typically defined
without reference to an underlying (pre-deformation) Calabi-Yau
manifold, the picture that we adopt here is that one starts with an
ordinary Calabi-Yau compactification, and then deforms that
compactification as parametrized by the fluxes and torsion.  However, 
it is not known that good compact examples of the types we discuss
are related by any physical process (such as domain walls
\refs{\GukovYA,\KachruNS}) or sensible mathematical deformation to a
Calabi-Yau background with D-branes and orientifolds.
For noncompact local models, however, one can make arbitrarily
small, continous deformations of the flux and torsion classes, and
we will to a large degree focus on such models here.  (Our observations
about the relationship to nongeometric models also holds for compact examples,
but the specific arguments there do not depend on any relation to a nearby Calabi-Yau).  
At any rate, we hope that the current work, combined with the many results
regarding the action of mirror symmetry
on D-branes, will be a useful guide to the story for fully compact models.

The fact that the compactifications we study are globally nongeometric
demonstrates that supergravity is insufficient.  Furthermore, the auxiliary fields for
hypermultiplets in type IIB induce superpotentials for the K\"ahler
moduli of the underlying Calabi-Yau geometry; at the minimum of the
resulting potential, the volumes of some cycles will generically be
string scale.  In \S6, we provide direct
arguments that worldsheet instanton effects are important.
The reader may sensibly object that most of our analysis
nonetheless uses the supergravity approximation.  To the extent that
we study {\it local\/} noncompact models, and can consider the
hypermultiplet auxiliary fields to parametrize small, continuous
deformations of Calabi-Yau backgrounds (as opposed to discrete
deformations), we are on good footing. Compact models will require a
stringy version of the mathematics of $SU(3)\times SU(3)$ structure, as well as need to satisfy additional constraints, as in \KashaniPoorSI.
Hopefully, the worldsheet perspective in this paper will provide the
first step in this direction.  More generally, we feel that our
worldsheet perspective gives a useful organization of and insight into
generalized geometries.

\newsec{Review}

In this section, we review some facts about compactifications that give
$\CN=2$ effective actions; this means that the Lagrangian is invariant
under off-shell $\CN=2$ supersymmetry transformations, but 
expectation values for auxiliary fields
break the $\CN=2$ supersymmetry.  We open in \S2.1\ with a discussion
of the $\CN=2$ superspace expansion
of \refs{\GrimmXP\deWitPQ\deRooMM-\BerkovitsCB}
for \thypermultiplets, and discuss the auxiliary field structure.
In \S2.2\ we review the relationship between spacetime supersymmetry
in four dimensions and $G$-structures on the compactification
manifolds. In \S2.3\ we review the relationship between spacetime and
worldsheet supersymmetry.  Finally, in \S2.4\ and \S2.5 we review some
basic facts about $SU(3)$ structure and $SU(3)\times SU(3)$ structure,
respectively.

\subsec{$\CN=2$ superspace expansion of \thypermultiplets}

In this paper, we study type IIB theories with closed string modes
that lie in $\CN=2$ supermultiplets.  As described in the previous
section, this makes sense in models for which $\CN=2$ supersymmetry is
broken to $\CN=1$ at a lower scale than the compactification scale, or
is so broken by local defects in the compactification.  Many of the
type II flux compactification models that have dominated the recent
literature on string model-building fall into this
class \refs{\DeWolfeNN,\KachruEM}, as do Ho\v{r}ava-Witten
compactifications \refs{\HoravaQA,\HoravaMA}\ for a broad range of
parameters consistent with coupling constant
unification \refs{\BanksDH,\BanksAY}. We focus on type IIB models in
this paper.

Following Ref.~\GirardelloWZ, our interest is in breaking $\CN=2$
supersymmetry to $\CN=1$ or $\CN=0$ through expectation values of
bosonic auxiliary fields in $\CN=2$ multiplets.  These expectation
values appear in nonzero (non total derivative) supersymmetry
transformations of the fermions, so that the state breaks
supersymmetry.  Note that consistent $\CN=1$ or $\CN=0$
compactifications to four dimensions include orientifolds which may
project out half or all of an $\CN=2$ multiplet. However, if $\CN=1$
supersymmetry survives to low energies, the surviving closed string
fields should descend from the original $\CN=2$ theory via the 
orientifold projection, and the surviving auxiliary
fields in the $\CN=1$ multiplets control the SUSY-breaking terms of this 
more realistic model.

In \refs{\LawrenceZK,\LawrenceKJ}, the focus was on vector multiplets
of the underlying type IIB $d=4$, $\CN=2$ model, and on hypermultiplets of type IIA. In this work we
focus on the hypermultiplets of type IIB (our results will also give the NS-NS auxiliary
fields for the type IIA vector multiplets).  There are various off-shell
extensions of $\CN=2$ multiplets whose on-shell bosons all have spin
zero.  However, there is a particular off-shell extension that appears
to be natural from the point of view of the string worldsheet.  It
follows from the $\CN=2$ superspace formalism of
Refs.~\refs{\GrimmXP,\deWitPQ,\deRooMM,\BerkovitsCB}.  In this
formalism, the anticommuting superspace coordinates are a pair of
spinor-valued Grassmann variables
$(\theta_{\alpha},\hat{\theta}_{\alpha})$ and their complex conjugates
$(\bar{\theta}_{\dot{\alpha}},\hat{\bar{\theta}}_{\dot{\alpha}})$.
Each pair is a doublet of Weyl spinors under the $SU(2)_R$ symmetry of
$d=4$, $\CN=2$ supersymmetry.  If we choose a direction in the doublet
representation of $SU(2)_R$, the corresponding Weyl spinor is the
superspace Grassmann variable of the $\CN=1$ subalgebra of the $\CN=2$
supersymmetry.

The doublet of Grassmann variables arises very naturally from the
worldsheet \refs{\LawrenceZK,\LawrenceKJ,\BerkovitsCB}.  For type II
strings with $\CN=(2,2)$ worldsheet supersymmetry, currents for
spacetime supersymmetry can be constructed from both the left- and
right-moving sectors of the worldsheet \refs{\BanksCY,\BanksYZ},
leading to a natural decomposition of $\CN=2$ spacetime supersymmetry
into two $\CN=1$ subalgebras.  The spacetime supersymmetries formed
from the left- and right-moving sectors of the worldsheet form an
$SU(2)_R$ doublet; the $SU(2)_R$ symmetry is nonperturbative on the
worldsheet.  As in \refs{\BerkovitsCB}, we take $\theta$ to be the
superspace variable corresponding to the $\CN=1$ subgroup of the
spacetime supersymmetry built from the left-moving sector of the
worldsheet, and $\hat{\theta}$ to be the superspace variable
corresponding to the $\CN=1$ subgroup of the spacetime supersymmetry
built from right-moving worldsheet fields.

The $\CN=2$ superfield for a hypermultiplet is chiral with respect
to the left moving supersymmetry and anti-chiral with respect to the
right-moving supersymmetry. Its expansion in the superspace Grassman
variables is:
\eqn\twohyper{
\eqalign{
   \CH^a & = w^a + \theta^\alpha \chi_\alpha^a
   + \bhthet^{\dot{\beta}} \hat{\bar{\chi}}_{\dot{\beta}}^a
   + \theta^2 y^a + \bhthet^2 \hat{\bar{y}}^a \cr
   \ \ & \ \ + \theta^\alpha \bhthet^{\dot{\beta}}
   \sigma^\mu_{\alpha\dot{\beta}} F_\mu^a  + \theta^\alpha \bhthet^2 \eta^a_{\alpha}
   + \bhthet^{\dot{\beta}}\theta^2
   \hat{\bar{\eta}}^a_{\dot{\beta}}  + \theta^2 \bhthet^2 C^a\ .
}}
Here $w^a$ is a complex scalar, and $F_{\mu} = \p_{\mu} \varphi^a$,
where $\varphi^a$ is also a complex scalar; $y^a$, $\hat{ \bar{ y}}^a$
and $C^a$ are auxiliary fields, and $a$ simply labels the moduli and
runs over the appropriate range of values, i.e., 
\eqn\arange{\eqalign{a &=1, \dots, h^{(1,1)}\qquad\hbox{type IIB,}\cr
           a &= 1, \dots, h^{(1,2)}\qquad\hbox{type IIA,}}}
in a Calabi-Yau compactification.

\subsubsec{\Thypermultiplets\ in type IIB}

Consider type IIB string theory compactified on a Calabi-Yau manifold
$X$, with $\{\omega_a\}$ a basis of $H^{(1,1)}(X)$.  The K\"ahler form
can be written as $J = t^a \omega_a$, where the $t^a$ are the real
Kahler moduli of the compactification.  Vacuum expectation values $B =
b^a \omega_a$ for the \NSNS\ 2-form potential also preserve spacetime
supersymmetry.  In this case we define complexified K\"ahler moduli
$w^a = b^a + i t^a$, which are again purely \NSNS\ fields.
Supersymmetry transformations act with a half unit of spectral flow,
mapping the NS sector to the R sector and vice-versa.  Therefore, the
field, $\varphi^a$ in Eq.~\twohyper\ is an RR field, and $y^a$ is
an \NSNS\ auxiliary field.  The field $\varphi^a$ can be constructed
as follows.  The RR two-form $C^{(2)}$ contributes massless modes via
the decomposition $C = \sum_a c^a \omega_a$.  The RR four-form
$C^{(4)}$ gives four-dimensional two-form potentials dual to scalars,
via the expansion $C^{(4)} = \sum_a \tilde{c}^a_{\mu\nu}\omega_a$.  We
write $\p_{\mu}\varphi^a = \p_{\mu} c^a + i ({}^*d \tilde{c}^a)_\mu$.

The universal hypermultiplet, which includes the four-dimensional
dilaton $\phi$ (not to be confused with the RR scalars $\varphi^a$
just discussed), arises in a different way.  For this multiplet, the
natural complex \NSNS\ scalar is written $w^{\phi} = a + i
e^{-2\phi}$, where the pseudoscalar $a$ is the four-dimensional dual of the \NSNS\
2-form $B_{\mu\nu}$.  However, the worldsheet naturally couples to
$B_{\mu\nu}$, not $a$.  Correspondingly, in Ref.~\refs{\BerkovitsCB},
the dilaton has a different superfield description than that described
in Eq.~\twohyper: one decomposes a real scalar $\CN=2$ superfield into
superfields for the graviton and dilaton multiplets, and the \NSNS\
2-form arises directly in the latter.  The corresponding auxiliary fields are
Ramond-Ramond.  However, from a four- or ten-dimensional 
spacetime point of view, there is no problem in principle with
writing the dilaton multiplet in the form \twohyper\ with $w^{\phi}$
as described, even if there is no obvious vertex operator description
of $w^\phi$.  We will find a thus natural candidate for $y^{\phi}$ from a
spacetime rather than worldsheet point of view. It would be interesting
and important to construct the auxiliary fields in the dilaton
multiplet as presented in \refs{\BerkovitsCB}.  It may also be important
to understand the auxiliary fields in other off-shell presentations of the
hypermultiplets, such as the various multiplets that arise from projective
superspace \refs{\LindstromKS}.  In particular, some string compactifications
in the literature -- {\it e.g.}\ 
\refs{\GranaNQ,\CamaraKU} -- have F-terms in the dilaton hypermultiplets which are
combinations of NS-NS and R-R fields, as we will discuss below.

In \refs{\BerkovitsCB}, the number of off-shell degrees of freedom in
the multiplet \twohyper\ is reduced by imposing additional reality
conditions $\del^2 \CH = \hat{\del}^2 \CH = 0$, where
$\del,\hat{\del}$ are defined in \refs{\BerkovitsCB}.  We find that
these constrain the components of $\CH$ such that $C^* \propto \p^2
w$, and $\hat{\bar y} = y^*$ (which is slightly different from the
condition written in \refs{\BerkovitsCB}).  In the discussion below,
we will not impose such conditions, so that $y$ and $\hat{\bar{y}}$
are independent.  This is consistent with the type IIA picture
discussed in \refs{\LawrenceZK}.  For example, a background with $y =
0$ and $\hat{\bar{y}} \neq 0$ corresponds to a background with an
$\CN=1$ supersymmetry preserved (with Grassmann superspace variables
$\theta$).  A noncompact example is the solution
of \refs{\MaldacenaYY,\ChamseddineMC}, as discussed
in~\refs{\LawrenceZK}.

\subsec{Spacetime supersymmetry and $G$-structures}

In ten dimensions, type IIB string theory contains two supercharges
$Q_N$ of the same chirality.  The supersymmetry transformations are
parametrized by two ten-dimensional positive-chirality Majorana-Weyl spinors $\eps_{N}$,
where $N=1,2$. For compactifications to four dimensions, we write
(\cf~Ref.~\refs{\GranaNY})
\eqn\mwdecompose{
	\eps_N = \zeta_{N+}\otimes\eta^N_{-} + \zeta_{N-}\otimes \eta^N_{+},
}
where $\zeta_{N\pm}$ are four dimensional spinors, $\eta^N_\pm$ six
dimensional spinors, and the subscripts~$\pm$ denote four dimensional
and six dimensional chirality, respectively.
\foot{The assigment of 6d chirality follows from the definition
of the four-dimensional chirality operator as $\gamma^5 = i \gamma^0\ldots \gamma^3$ and
the 6d chirality operator as in the Appendix.  Using the definitions of the
10d Clifford algebra in eq. (2.2) of \GranaNY, then
$\Gamma^{11} = \Gamma^0 \ldots \Gamma^9 = - \gamma^5_{(4d)}\otimes \gamma^7_{(6d)}$.}

\subsubsec{$\CN=2$ supersymmetry and $SU(3)\times SU(3)$ structures.}

We begin by considering a compactification which is locally a smooth six
dimensional manifold $M$, which is well described by
supergravity.\foot{However, we will point out in \S3, \S6\ that the vacua in
truly compact $SU(3) \times SU(3)$ structure models will generically
have string-scale features; this is one of many dangerous games we
play in this paper.}  Following \refs{\GranaNY,\GranaHR}, we demand
that the full effective action (including all of the massive
Kaluza-Klein and string modes) be invariant under $\CN=2$
supersymmetry.  Note that this condition is compatible with the
presence of nonvanishing expectation values of auxiliary fields:  
the {\it action} is still invariant under $\CN=2$
supersymmetry, but the state with these
expectation values is not. Expectation values for
the auxiliary fields break $\CN=2$ to $\CN=1$ or
$\CN=0$.

For the action of $\CN=2$ supersymmetry to be well-defined, the
spinors $\eta^N$ must be globally well-defined.  When the solution is
smooth and reliably described by supergravity, one typically demands,
as in \refs{\GranaNY,\GranaHR}, that the spinors are also
nowhere-vanishing.  This condition usually follows from the demand
that for a {\it supersymmetric}\ background, the spinor $\eta^N$ be
covariantly constant, which implies that its norm is constant.

When the supersymmetry is nonlinearly realized, and still described by
a nowhere-vanishing spinor, it is possible for the corresponding
$\CN=2$ supersymmetry to be broken at a low scale compared to the
Kaluza-Klein scale.  Spinors that {\it do} vanish at points or at loci
of finite codimension cannot be covariantly constant, and in
nonsingular geometric compactifications correspond to a broken
supersymmetry in the set of local ten-dimensional supersymmetries.  
We expect the energy scale of breaking to generically
be the Kaluza-Klein scale.
\foot{We thank D. Waldram and especially M. Gra\~{n}a for patient
correspondence on these points.}

In the work described here, we have in mind the case that the $\CN=2$
supersymmetry is broken by expectation values of auxiliary fields at a
low scale compared to the Kaluza-Klein scale.  We therefore consider
backgrounds with two nowhere-vanishing spinors.  For a spinor to be
globally well-defined, it must be invariant under the structure group
$G \subset \Spin(6)$ of the spinor bundle on $M$.  Thus, the
decomposition of the ${\bf 4}$ of $\Spin(6)$ into irreducible
representations of $G$ must contain a singlet. The generic ({\it i.e.}
largest) structure group $G$ with these properties is $SU(3)$.  Each
invariant, nowhere-vanishing spinor $\eta^N$ defines an $SU(3)$
structure on the spinor bundle.  (This structure group is inherited by
the frame bundle\foot{The frame bundle is defined by its sections: a
local section of the frame bundle is a choice of vielbein basis, i.e.,
a ``frame'' of six 1-forms (or vectors) on each open set $U\in M$.
Usually, there is no global section; in the special case that one
exists, there are six global 1-forms $e^A$, and $M$ is said to be
parallizable.} on $M$, so it is also possible to describe an $SU(3)$
structure in terms of the frame bundle without reference to spinors.)

Since the $\Spin(6)$ spinors $\eta^1_{\pm}$ and $\eta^2_{\pm}$ of
Eq.~\mwdecompose\ need not be proportional to one another, each
generically defines an $SU(3)$ structure distinct from the other.
Thus, we have two structures, $SU(3)_N$, for $N=1,2$.  This is natural
from the point of view of the worldsheet, as we discuss further below:
the spinors $\eta^1$ and $\eta^2$ are generated from the left- and
right-moving sectors of the worldsheet, respectively.  Each chiral
sector of the worldsheet has its own associated $SU(3)$ structure.

It is possible to combine the two $SU(3)_N$ structures into a single
$SU(3)\times SU(3)$ structure in the ``generalized complex geometry''
of Hitchin et al~\refs{\GranaNY,\HitchinUT,\GualtieriThesis}.  In this
case, the generalized tangent space of interest is (a bundle extention
of) $T\oplus T^*(M)$, with structure group $SU(3)\times SU(3)$.  In
this generalized geometry, the role played by the $SU(3)_N$ invariant
$\Spin(6)$ spinors $\eta^N_\pm$ in the previous discussion is now
played by $SU(3)_1\times SU(3)_2$ invariant $\Spin(6,6)$ pure spinors:
$\Omega_+\propto \re\eta^1_+\otimes\eta^2_+$ of positive chirality and
$\Omega_-\propto\re\eta^1_+\otimes\eta^2_-$ of negative chirality.  In
this way of writing the pure spinors, $SU(3)_1$ acts on the left and
$SU(3)_2$ acts on the right.  To reproduce the earlier discussion, all
that is needed is a projection from $T\oplus T^*(M)$ to $T(M)$.  As
described in Ref.~\GualtieriThesis, there are two canonical choices of
this projection.  One gives the group $SU(3)_1$ associated to
$\eta^1_\pm$, and the other gives the group $SU(3)_2$ associated to
$\eta^2_\pm$.

These various ways of encoding $SU(3)\times SU(3)$ structure are
closely related to the various presentations of the ``doubled torus"
in \refs{\HullIN} used to describe stringy torus fibrations. (See, for example,
\refs{\LawrenceMA}\ for a systematic discussion of this formalism.)
In that work, one replaces a $T^n$ factor (or fiber) in the target
space with $T^{2n}$, on which the T-duality group acts linearly. One
may choose a polarization that splits this torus into two
$n$-dimensional factors. One choice is to split them into two $T^n$
factors described by left- or right-moving chiral bosons on the
worldsheet.  This is analogous to the tack we will take in this paper.
Alternatively, one may split the doubled torus into a direct sum of
the original torus and its dual.  This is closer in spirit to the
discussion in \refs{\HitchinUT,\GualtieriThesis}.

\subsubsec{$\CN=4$ supersymmetry and local versus global $SU(2)$ structures}

For a generic $SU(3)\times SU(3)$ structure, the spinors are locally
independent, and are only parallel at isolated points.  If they are
{\it never} parallel, then the two spinors define an $SU(2)$
structure. This allows one to define an $\CN=4$ supersymmetry acting on
the four-dimensional theory, by reducing each of the ten-dimensional
spinors $\eps_{1,2}$ on either of the six-dimensional spinors
$\eta_{1,2}$.

If the spinors are parallel at points, there is a local but not a
global $SU(2)$ structure.  In principle, one could still reduce each
of $\eps_{1,2}$ on either of $\eta_{1,2}$ and so define an $\CN=4$
supersymmetry.  However, the fact that the spinors $\eta_{1,2}$
coincide at points means that one of the putative supercharges in the
$\CN=4$ algebra will come from a reduction on a spinor which vanishes at
specific points in the moduli space.  As discussed above, this means
that the $\CN=4$ supersymmetry will be broken to $N \leq 2$, generically
at the Kaluza-Klein scale.

\subsec{Worldsheet vs. spacetime supersymmetry}

$\CN=2$ spacetime supersymmetry in four dimensions requires $\CN=(2,2)$
{\it worldsheet} supersymmetry for the $c = 9$ superconformal field
theory describing the compact CFT \refs{\BanksCY,\BanksYZ}.  The
$SU(2)_R$ doublet of supercharges in the $(-1/2,-1/2)$ picture can be
written as:
\eqn\sutwodoublet{
	\left( \matrix{ Q_{L,\alpha}(z) \cr \hat{Q}_{R,\beta}(\bar{z}) }\right)
	= \left(\matrix{e^{-\phi_L/2} S_{\alpha,L} \bar{\Sigma}_L(z) \cr
	e^{-\phi_R/2} S_{\beta,R} \bar{\Sigma}_R(\bar{z})} \right)
}
Here $\phi_{L,R}$ come from the bosonization of the superconformal
ghosts, $S_{\alpha}$ are the spin fields for the $\QR^4$ sigma model
factor of the CFT, and $\bar{\Sigma}_{L,R}$ are the $U(1)$ charge $-\frac{3}{2}$ 
spectral flow operators for the $c=9$ compact SCFT, mapping NS $\leftrightarrow$ R.  If the
compact CFT is a sigma model, then $\Sigma_{L,R}$ can be written as
spin fields for the sigma model coordinates, and transform in the
spinor representation of $\Spin(6)$. Supersymmetry requires that this
be a singlet of $SU(3)\subset SO(6)$.  Thus there is a map between the
spectral flow operators and these spinors.

The standard example of a sigma model with $\CN=(2,2)$ worldsheet
supersymmetry is one with a Calabi-Yau target space $M$.  In this case
$\eta^1_{\pm} = \eta^2_{\pm} = \eta_{\pm}$ and $\Del_m \eta_{\pm} = 0$.  The
Levi-Civita connection on $M$ has $SU(3)$ {\it holonomy}, and thus $M$ is guaranteed to be
Ricci-flat and K\"ahler. This is not the most general
possibility for $\CN=(2,2)$ sigma models \refs{\GatesNK}. In the
presence of nonvanishing \NSNS\ three-form flux, $\CN=(2,2)$
supersymmetry is preserved if there are two {\it almost} complex
structures $J_{\pm}$ such that
\eqn\bihermite{
	\Del_{\mu} J_{\pm}^{\nu}{}_{\lambda} \mp 
	\half \left(H^{\nu}_{\mu\rho} J^{\rho}_{\pm \lambda} - 
	H^{\rho}_{\mu\lambda} J^{\nu}_\pm {}_{\rho} \right)= 0.
}
Here, $J_+$ and $J_-$ should be identified with $J^1$ and $J^2$ of
Sec.~2.  If $\psi^{\mu}_L(z)$, $\psi^\mu_R(\bz)$ are the left- and
right-moving spacetime fermions polarized along $M$, then $J_{L,R} =
J_{L,R,\mu\nu}\psi^{\mu}_{L,R}\psi^{\nu}_{L,R}$ are the left and right
moving worldsheet $U(1)_R$ currents in the $\CN=(2,2)$ algebra. One
may construct $\Sigma_{L,R}$ by bosonizing these $U(1)$ currents, and
these will be mapped to spinors which satisfy
\eqn\spinorswithtorsion{
	\Del_{\mu} \eta \pm \frac{1}{8} H_{\mu\nu\rho} \Gamma^{\nu\rho} \eta = 0,
}
where the $\pm$ is correlated with the $d=10$, $\CN=2$ supersymmetry
from which $\eta^N$ descends (i.e., $+\leftrightarrow\eta^1$ and
$-\leftrightarrow\eta^2$). These backgrounds correspond to a
particular class of $SU(3)\times SU(3)$ structure compactifications. 
We are working with an off-shell presentation of the hypermultiplets \twohyper\ for which the
auxiliary fields are NS-NS.  Other presentations may involve RR fields.  The RR flux would modify 
\bihermite\ and \spinorswithtorsion, and may break additional supersymmetry. However, since we do not know how to treat RR backgrounds in the RNS worldsheet formalism, we will not study these effects in this section (see \amiretal). 

We will be particularly interested in cases where $\CN=2$ spacetime
supersymmetry is broken.  If the supersymmetry is broken by \NSNS\
deformations to $\CN=1$ and the dilaton does not become too large, the
worldsheet supersymmetry is generically broken to $\CN=(2,1)$; if
spacetime supersymmetry is broken entirely then the worldsheet
supersymmetry is broken to $\CN=(1,1)$.
It is also possible that
supersymmetry is broken simply because the physical states no longer
satisfy the $R$-charge quantization rule described
in \refs{\BanksCY,\BanksYZ}.  We believe that in terms of the $\CN=1$ spacetime
supersymmetry associated with this $R$-symmetry, this breaking will be
through D-terms, as is true in the open string case \refs{\DouglasGI}. Following that
work, we expect the argument to run like this: the complex NS-NS scalars are described by
a vertex operator with $U(1)_R$ charge, and so a change of the R-charge
affects equally both real scalars in the spacetime multiplet.
In particular the mass shifts will be the same for both.  This is characteristic of D-term breaking;
F-term breaking leads to mass splittings between scalars in chiral multiplets.
The D-terms are auxiliary fields in the vector multiplets. 
We leave verification of this scenario for future work. 
Meanwhile, the statement that auxiliary fields in hypermultiplets break some 
of the $\CN=(2,2)$ worldsheet supersymmetry is fully consistent with the results in this paper.
In particular this typically means that one or both of the $U(1)_R$ symmetries are broken.

The worldsheet manifestation of this is as follows.  When spacetime
supersymmetry is broken through expectation values for the auxiliary
fields $y,\hat{\bar{y}}$, the $\CN=2$ transformations will still act
on the fields, albeit nonlinearly.  On the worldsheet, we will find
that the operators corresponding to $y,\hat{\bar{y}}$ explicitly break
the $U(1)_R$ charge.  The $R$-current exists but no longer
satisfies \bihermite; as we will describe below, the left hand side
will contain torsion terms.

\subsec{Review of $SU(3)$ structure}

The most general case that we are interested in has two $SU(3)$
structures.  To understand them it is helpful to focus first on one
$SU(3)$ structure---this will also describe the well-studied case in
which the two $SU(3)$ structures are parallel and can be made
identical.

Manifolds with $SU(3)$ structure can be classified by a set of
intrinsic torsion classes.  These encode the failure of the
corresponding positive and negative chirality spinors $\eta_{\pm}$ to
be covariantly constant with respect to the Levi-Civita connection:
\eqn\spinorscovder{
\nabla_m \eta_\pm =  (q_m + i \tilde q_m \gamma_7)\eta_\pm + i q_{mn} \gamma^n \eta_\mp ,
}
where $\gamma_7$ is the six-dimensional chirality operator, and
$\tilde q_m, q_m, q_{mn}$ are determined by the intrinsic torsion of
the manifold \GranaSN.

Alternatively, one may define an almost complex structure
via:\foot{See Appendix~A for a complete discussion of our
normalization conventions.}
\eqn\suthreeacs{
	J_{mn} = - i \bar \eta_{\pm} \gamma_{mn} \gamma_7 \eta_{\pm},
}
where $\eta^{\dagger}_{\pm} \eta_{\pm} = 1$.\foot{Note that our
definitions and normalizations differ by factors of 2 from those given
in \refs{\GranaNY,\GranaHR}.  Our definitions are consistent with the
conventions given in Appendix~A.  In particular there is a factor of 2
difference that appears in the Fierz identity given in Appendix A.}
The torsion classes measure the failure of $J$ to be covariantly
conserved.  A third description \refs{\ChiossiSal} is as follows:
Define the two-form $J$ with coefficients $J_{mn}$; this has index
structure $(1,1)$ with respect to the almost complex structure.
Define also the $(3,0)$ form $\Omega$ with coefficients
\eqn\suthreetopform{
	\Omega_{mnp} = - i \bar \eta_- \gamma_{mnp} \eta_+
}
The torsion classes, which measure the deviation of the $SU(3)$
structure manifold from having (Levi-Civita) $SU(3)$ holonomy, can
then be defined as:
\eqn\djdom{
\eqalign{
dJ & =  - {3 \over 2} {\rm Im} (W_1 \overline \Omega) + W_4 \wedge J + W_3 \cr
d\Omega & =  W_1 J^2 + W_2 \wedge J + \overline W_5 \wedge \Omega\ .
}}
Here $W_1$ is a complex 0-form, $W_2$ is a complex (1,1) form where
$W_2 \wedge J$ is primitive with respect to $J_{mn}$, $W_3$ is a real
primitive $(2,1) \oplus (1,2)$ form,\foot{A form $\omega$ is primitive
with respect to $J$ if $\omega\wedge J = 0$.} $W_4$ is a real
one-form, and $W_5$ is a $(1,0)$ form.  Note also that $dJ$ can
include a $(3,0)\oplus (0,3)$ piece in addition to a $(1,2)\oplus
(2,1)$ component because the almost complex structure is generically
not integrable.  Similarly, $d\Omega$ can include $(2,2)$
components. Using the Fierz identities given in Appendix~A, we can
define $q_m$, $\tilde q_m$ and $q_{mn}$ in terms of the $W_i$.

Each of the $W_i$ lives in a definite representation of the $SU(3)$
structure group.  Any given representation is most easily found by
noting that holomorphic indices (with respect to the almost complex
structure $J_m{}^n$) lie in the ${\bf 3}$ of $SU(3)$, while
antiholomorphic indices lie in the ${\bf \bar{3}}$ of $SU(3)$.  Thus,
$W_1$ is a complex $SU(3)$ singlet; $W_2$ is a complex form in the
${\bf 8}$ of $SU(3)$; $W_3$ in the ${\bf 6}\oplus {\bf \bar{6}}$ of
$SU(3)$; and $W_4, W_5$ lie in the ${\bf 3} \oplus {\bf\bar{3}}$ of
$SU(3)$.

Following \refs{\GranaSN,\GranaSV,\GranaBG,\FidanzaZI}, we can
similarly expand the 3-form $H$ in this $(J,\Omega)$ basis as
\eqn\hexp{
H  =  - {3 \over 2} {\rm Im} (H_1 \overline \Omega) + H_3 + H_4 \wedge J, 
}
and we will find it useful to do so in the following sections. The
$H_k$ lie in the same representations as $W_k$, for $k=1,3,4$.

\subsubsec{Intrinsic torsion and the spin connection}

Our computations in \S3\ will use the relationship between the
intrinsic torsion and the components of the spin connection decomposed
according to the almost complex structure.

Given a vielbein $\{e^A\}$, $A = 1,\dots,6$, considered as a
collection of one-forms, we define a complex vielbein
\eqn\viels{
\eqalign{
e^a &= e^{A = 2a-1} + i e^{A = 2a} \cr
e^{\bar a} &= e^{A = 2a-1} - i e^{A = 2a},
}}
where $a=1,2,3$.  In this basis \refs{\ChiossiSal}\
\eqn\joviel{
\eqalign{
J &=i g_{a \bar a}e^a \wedge e^{\bar a} \cr
\Omega &= e^1 \wedge e^2 \wedge e^3,
}}
where $g_{a\bar{a}} = \half \eta_{a\bar{a}} = \half \delta_{a \bar a}$
is the flat metric in complex coordinates.  The 2-form $J$ defines an
almost complex structure after raising one index with the inverse
metric $g^{\bar a a}$.

Using the Cartan structure equations,
\eqn\sc{
\eqalign{
de^a & = - \omega^a{}_c \wedge e^c - \omega^a{}_{\bar{c}} \wedge
e^{\bar{c}}\cr & = - \omega_b{}^a{}_c e^b \wedge e^c
- \omega_{\bar{b}}{}^a{}_c e^{\bar{b}}\wedge e^c -
\omega_b{}^a{}_{\bar{c}} e^b \wedge e^{\bar{c}} - \omega_{\bar{b}}{}^a{}_{\bar{c}} e^{\bar{b}}
\wedge e^{\bar{c}}
}}
combined with \djdom\ and \joviel, we can relate components of the
spin connection with specific complex indices to different intrinsic
torsion classes.  Specifically, we find that
\eqn\scwone{
\eqalign{
W_1 &= {4 \over 3} \epsilon^{\bar a \bar b \bar c} \omega_{\bar a \bar b \bar c}\ ;
\ \ \ \ \ H_1 = - \frac{2i}{9}  \epsilon^{\bar a \bar b \bar c} H_{\bar a \bar b \bar c}\cr
(W_2)_{a \bar b} & =  i \Omega^{\bar{c}\bar{d}}{}_a \omega_{\bar{c}\bar{d}\bar{b}}
	- \frac{i}{3} g_{a\bar{b}} \Omega^{\bar{c}\bar{d}\bar{f}}\omega_{\bar{c}\bar{d}\bar{f}}\ .
}}
(See also \refs{\CamaraCZ}.)
Note in particular that the totally antisymmetric part of $\omega$
will not contribte to $W_2$.

\subsec{Review of $SU(3)\times SU(3)$ structure}

We will find that for general values of the hypermultiplet auxiliary
fields $y$ and $\hat{\bar{y}}$, the $\Spin(6)$ spinors $\eta^N$
($N=1,2$) are not parallel, so that we have two distinct $SU(3)$
structures.  Locally on $M$, the two spinors are independent almost everywhere, 
and may become parallel only at isolated points.  In a
neighborhood in which the two spinors are everywhere independent, the
intersection $SU(2) = SU(3)_1 \cap SU(3)_2$ defines a local $SU(2)$
structure.  However, this intersection is enlarged to $SU(3)$ at the
special points where the two spinors $\eta^{N}$ become parallel.
Therefore, globally, there are two $SU(3)$ structures, but there is no
global $SU(2)$ structure.

Given the doublet $\eta^N$, one may define a doublet of real 2-forms
$J^N$ and a doublet of complex three forms $\Omega^N$,
\eqn\JsandOmegas{
J^{N}_{mn} = - {i\over 2} \bar \eta^{N} \gamma_{mn} \gamma_7 \eta^{N} \quad\hbox{and}\quad \Omega^{N}_{mnp} = - \frac{i}{2} \bar \eta^{N} \gamma_{mnp} (1+\gamma_7) \eta^{N} .
}
Here, the pair $(J^N,\Omega^N)$ provides an equivalent definition of
the $SU(3)_N$ structure.  Accordingly we have a doublet
$W^N_1, \ldots, W^N_5$ of the five torsion classes defined in
Eq.~\djdom.  We will argue that $N=1,2$ correspond to $y,\hat{\bar
y}$, respectively. Furthermore, for each $SU(3)$ structure, there is
an almost complex structure with respect to which we can write a
vielbein with complex tangent frame indices.  For each such almost
complex structure, the intrinsic torsion classes $W^N_k$ can be
written in terms of the spin connection as in \scwone.

Instead of defining a doublet of $SU(3)_N$ torsion classes, one could
instead define a single set of $SU(3)\times SU(3)$ torsion classes;
typically, one would define them in terms of the pure spinors built
from $\eta^N_{\pm}$, as in \refs{\GranaSN}. However, the two
descriptions are equivalent, and the formulation given here will be
the most useful for our purposes.

\newsec{Worldsheet vertex operators for auxiliary fields}

In this section we will follow \refs{\LawrenceZK,\LawrenceKJ} and use
worldsheet techniques to compute $y,\hat{\bar{y}}$ for K\"ahler moduli
in Calabi-Yau compactifications of type IIB string theory.

\subsec{Worldsheet supersymmetry and target space geometry}

Expectation values $y \neq 0$ or $\hat{\bar{y}} \neq 0$ break the left- or
right-moving $\CN=2$ superconformal symmetries, respectively, down to
$\CN=1$.
In particular, $\hat{\bar{y}} \neq 0$, $y = 0$, corresponds to a
background with $\CN=(2,1)$ supersymmetry.  This observation will be
useful in interpreting the vertex operator calculation.

A general $\CN=(1,1)$ worldsheet sigma model with \NSNS\ $H$-flux has
the fermion bilinear terms\foot{See Appendix A for our conventions for
$H$.}
\eqn\wsl{
\eqalign{
{\cal L} = & -ig_{\mu \nu} \psi^\mu_+ {\cal D}_- \psi^\nu_+ - i g_{\mu\nu}
	\psi^{\mu}_- \CD_+ \psi^{\nu}_-\cr
g_{\mu\nu}\psi^{\mu}_{\pm} \CD_{\mp} \psi^{\nu}_{\pm} \equiv & 
g_{\mu \nu} \psi^\mu_\pm \partial_\mp \psi^\nu_\pm +
(g_{\mu \nu}\Gamma^\nu_{\lambda \rho} \pm \half H_{\mu \lambda \rho}) \psi^\mu_\pm \partial_\mp X^\lambda \psi^\rho_\pm\,,
}}
where $\psi_-$ ($\psi_+$) are left- (right)-moving worldsheet fermions. 
In addition, there are four-fermion terms of the form
$R_\pm\psi\psi\psi\psi$ where $R_\pm$ is curvature built from the
connection $\Gamma^{\mu}_{\pm,\nu\rho}
= \Gamma^{\mu}_{\nu\rho} \mp \half H^{\mu}_{\nu\rho}$.  Worldsheet
$\CN=2$ supersymmetry for either the left or the right movers requires
the existence of a complex structure $J_-$ or $J_+$, respectively,
which satisfies \bihermite\ \refs{\HullZY}.  (If both $J_{\pm}$
satisfy \bihermite, then the theory has $\CN=(2,2)$
supersymmetry). This condition implies that the metric must be
hermitian, the \NSNS\ three-form must be of holomorphic type (2,1)
with respect to the almost complex structure $J_+$ or $J_-$, and
\eqn\hpmt{
H_{ij \bar k} = \pm T_{ij \bar k} \equiv \pm i (\partial_j g_{i \bar k} - \partial_i g_{j \bar k})\ ,
}
where the sign in \hpmt\ is correlated with the sign $\pm$ in
$J_{\pm}$.  (In the case of (2,2) supersymmetry, the geometry is
bihermitian.)

\subsec{Vertex operators for auxiliary fields}

Vertex operators for auxiliary fields $y,\hat{\bar{y}}$ are calculated
as follows \refs{\LawrenceZK,\LawrenceKJ,\AtickGY,\DineGJ}. For $y^a$,
we begin with the vertex operator for complexified K\"ahler modulus
$w^a = b^a+it^a$ in the $(-1,0)$ picture,
\eqn\vopgen{
	V_{w^a}^{(-1,0)} = e^{-\phi_-(z)} \CO_{\half, 1} 
}
where $\CO_{\half, 1}$ has left-moving conformal dimension $\Delta
= \half$ and right-moving conformal dimension $\Delta = 1$.  Let
$\Omega$ be the operator generating one unit of spectral flow (from NS
to NS) on the left-movers: for Calabi-Yau compactifications, it is
$\Omega = \Omega_{ijk}\psi^i_-\psi^j_-\psi^k_-$.  The vertex operator
for $y^a$ is found via the OPE:
\eqn\auxfieldope{
	\Omega(z) \CO_{\half,1}(u) \sim \frac{V_{y^a}}{(z - u)}
        + {\rm nonsingular}.
}
The vertex operator for $\hat{\bar{y}}$ is constructed identically by
exchanging left and right movers.  We are playing another dangerous
game in writing such vertex operators, as they generally do not
correspond to propagating modes.  Nonetheless,
following \refs{\AtickGY,\DineGJ}, they do appear as coefficients of
contact terms in certain OPEs, in precisely the fashion dictated by
supersymmetry.\foot{See \refs{\LawrenceBK} for a recent discussion of
this in the context of open string theory.}  Furthermore, the
identification of these fields with specific fluxes and torsion
classes on the manifold $M$ matches the spacetime arguments we will
provide in \S4\ and \S5.

K\"ahler moduli $w^a$ correspond to $(1,1)$ forms $\delta g^a_{i\jb}$.
The $(-1,0)$ picture vertex operator is:
\eqn\wvertex{
V_{w^a}^{(-1,0)} = e^{- \phi_-} \delta g^a_{\bar i  l} \psi_-^{\bar i} \partial_+ X^{ l}+e^{- \phi} \delta g^a_{i \bar l} \psi_-^{ i} \partial_+ X^{\bar l}.
}
In order to compute the OPEs, we will expand around a constant
background field $X$ in Riemannian normal coordinates, and pretend
that we are working at large radius.\foot{As noted above, this is very
dangerous.}  The scalars and fermions in this expansion have canonical
kinetic terms, making a perturbative calculation of the OPEs
straightforward.  Let the vielbein be $e_{\mu}^m$, with $m = 1,\ldots
6$ the frame indices, which can be organized into holomorphic and
antiholomorphic indices $a,\bar{a} = 1,\ldots 3$ by the almost complex
structure, as discussed in \S2.4\ above.  Following the discussion and
notation of \refs{\AlvarezGaumeHN}, we denote the components of the
bosonic fluctuation relative to the vielbein basis by $\xi^m$, with
fermionic superpartners $\psi^m = e^n_{\mu} \psi^{\mu}$. The quadratic
fermion kinetic terms \wsl\ become
\eqn\fkinframe{
\eqalign{
{\cal L} = & -i \eta_{mn} \psi^m_+ D_- \psi^n_+ - i \eta_{mn}
	\psi^m_- D_+ \psi^{n}_- \cr
	\eta_{mn}\psi^m_{\pm} D_{\mp} \psi^{n}_\pm & \equiv
	\psi^m_{\pm} \left[ \eta_{mn} \p_{\mp} \psi^{n}_\pm + 
	\left( \omega_{mnp} \pm \half H_{mnp} \right) \psi^n_{\pm} \p_{\mp} \xi^p \right ] \ ,
}}
where $\omega_{m}{}^n{}_{p} = e_m^{\mu} \omega_{\mu}{}^n{}_p$ is the
spin connection on $M$ with the 1-form index converted to the vielbein basis.

Let $\delta g_{i\bar{j}}$ be a deformation of the K\"ahler structure
of the metric, corresponding to the scalar $w$ in a \thypermultiplet.
The corresponding $(-1,0)$-picture vertex operator is
\eqn\wvertexframe{
	V_{w}^{(-1,0)} = e^{- \phi} \delta g_{\bar a b} 
        \psi_-^{\bar a} \partial_+ \xi^{b}+e^{- \phi} 
        \delta g_{a \bar b} \psi_-^{a} \partial_+ \xi^{\bar b}\ , }
where $\delta g_{a\bar b} \equiv\delta g_{i\jb}\, e_{a}^i
e_b^{\jb}$. Here and below, we have suppressed the upper index $a$ on
$w$ and $\delta g$ to avoid confusion with the complex vielbein
indices $a,\bar a$.

The spectral flow operator is:
\eqn\specflow{
\Omega(z) \equiv \frac{1}{3!}\Omega_{ijk}\psi_-^i \psi_-^j \psi_-^k = \frac{1}{6}
\eps_{abc} \psi_-^{a}\psi_-^{b}\psi_-^{c}.
}
Using the operator product expansions $\psi_-^{\bar
a} \psi_-^b \sim \eta^{\bar a b}$ and $\psi_-^{a} \psi_-^b \sim 0$, we
find that
\eqn\dvertex{
V_{y}^{(0,0)} = \delta g_{\bar a c} g^{\bar a d}
\Omega_{dab}\psi_-^{a}\psi_-^{b} \partial_+ \xi^{ c}, }
and similarly,
\eqn\dvertexright{
V_{\hat{\bar y}}^{(0,0)} =  \delta g_{a  \bar c} g^{a \bar d} 
\bar\Omega_{{\bar d}{\bar a}{\bar b}}
\psi_+^{{\bar a}}\psi_+^{{\bar b}} \partial_- \xi^{{\bar c}},
}
with similar expressions when $i\delta g_{i\bar\jmath}$ is replaced by
$\delta(B+ig)_{i\bar\jmath}$.  We interpret these vertex operators as
deformations of the Lagrangian \fkinframe.  Due to the appearance of
$\Omega_{dab}$ and $\bar\Omega_{\bar d\bar a\bar b}$ in the
expressions for $V_y$ and $V_{\hat{\bar y}}$, the quantity
$\omega \pm \half H$ in \fkinframe\ is deformed by purely $(3,0)$ and
$(0,3)$ components $\delta \omega \pm\half \delta H$.  Note that $H$
is automatically antisymmetric in all indices.  The definition of
$\omega_{abc}$ guarantees that it is antisymmetric in $b$ and $c$; the
fermion couplings in \fkinframe, \dvertex\ and \dvertexright\ are
consistent with this.

We have not performed the analogous computation for the universal
hypermultiplet $y_{\phi}$.  Instead, will deduce the corresponding auxiliary
field from spacetime considerations in Secs.~4 and 5.

\subsec{Target space interpretation}

Following \refs{\LawrenceZK,\LawrenceKJ}, our goal is to interpret the
results of the previous subsection in terms of known fluxes and target
space structures. We will begin by organizing the results according to
the amount of broken {\it worldsheet}\ supersymmetry. At the end of
this subsection we will then identify the auxiliary fields in terms of
intrinsic torsion classes for string backgrounds with local
$SU(3)\times SU(3)$ structure.

The basic results are as follows.  We wish to describe independent
deformations of the two auxiliary fields $y,\hat{\bar{y}}$.  In type
IIA compactifications, $y$ and $\hat{\bar y}$ can be independently
tuned, while staying within the class of manifolds with $SU(3)$ structure
and $H$-flux of the type $W_{3,4}$ in \djdom\ and $H_{3,4}$ in \hexp.
This is because the torsion and $H$-flux can be adjusted independently,
and the $H$-flux couples with opposite sign to left- and right-moving worldsheet fermions.

However, for type IIB compactifications, we will see that there are a
class of auxiliary fields for which the $H$-flux does not contribute;
the deformations are entirely geometric.  The fields $y,\hat{\bar{y}}$
must correspond to different geometric structures coupling to the left
and right movers. Each defines an $SU(3)$ structure.  Thus, we expect
that the compactifications with $y,\hat{\bar{y}} \neq 0$ are locally
manifolds with $SU(3)\times SU(3)$ structure.  On the other hand,
geometric deformations couple identically to left and right movers,
and the only way to enforce the statement that a given $SU(3)$
structure couples chirally is if the local patches of the
compactification are glued together with transformations which act
chirally. These transformations cannot be diffeomorphisms---they must
involve nontrivial reshufflings of string theoretic degrees of
freedom, as does T-duality.  Therefore, the generic manifold with
either or both of $y,\hat{\bar{y}}$ must be a nongeometric
compactification along the lines of 
\refs{\DabholkarSY, \DabholkarVE,\HullHK,\HellermanAX,\HullIN,\FlournoyVN}.

\subsubsec{$\N=(2,2)$ supersymmetry}

In case of $\N=(2,2)$ supersymmetry, we require that the spin
connection and \NSNS\ three-form, both with lowered vielbein indices,
have no $(3,0)$ or $(0,3)$ components.  This follows from the fact
that the torsion is a derivative of the metric $g$, with
$g \sim \partial \bar \partial K$ for a real potential $K$, as
required by supersymmetry. Therefore, if $\CN=(2,2)$ supersymmetry is
to be preserved, a deformation of the form \dvertex\ must be
accompanied by a change in the almost complex structure.

If the deformation contains no $H$ flux, the resulting background is a
Calabi-Yau background.  If the deformation has $H\neq 0$, then there
must be two almost complex structures satisfying \bihermite.  The
resulting six-dimensional background will no longer have $SU(3)$
holonomy, but instead an $SU(3)\times SU(3)$ structure.  As discussed
in \S2.3, the two almost complex structures satisfying \bihermite\
allow one to construct two conserved $U(1)_R$ charges on the
worldsheet, which are part of the $\CN=(2,2)$ superconformal
algebra.\foot{One could also attempt to construct two non-conserved
$U(1)_R$ currents, $J_{L,\mu\nu} \psi^{\mu}_R \psi^{\nu}_R$ and
$J_{R,\mu\nu} \psi^{\mu}_L \psi^{\nu}_L$.  These can be bosonized and
used to construct additional non-conserved spacetime supercharges in
an $N=4$ algebra.  We expect the corresponding supersymmetries to be
broken at the Kaluza-Klein scale.}

When $\CN=(2,2)$ supersymmetry is preserved, as long as the $U(1)_R$
charge of physical states remains appropriately quantized, we have not
broken the spacetime supersymmetry.  Given the assumption that
violating the R-charge quantization rule of \refs{\BanksCY,\BanksYZ}
corresponds to turning on auxiliary fields in vector multiplets, 
backgrounds corresponding to $y \neq 0$ and/or $\hat{\bar{y}} \neq 0$ must correspond to
backgrounds with reduced worldsheet supersymmetry.

\subsubsec{$\CN=(2,1)$ supersymmetry}

A background for which $y = 0$ and $\hat{\bar{y}} \neq 0$ (and the
auxiliary fields of the complex structure moduli vanish) will preserve
$\CN=1$ spacetime supersymmetry.  This requires $\CN=(2,1)$
supersymmetry on the worldsheet \refs{\BanksCY,\BanksYZ}.

The K\"ahler deformations and deformations of the auxiliary fields can
be classified according to their representations with respect to the
$SU(3)$ structure group.  The holomorphic and antiholomorphic indices
of tensors on the target space transform in the ${\bf 3}$ and
${\bf \bar{3}}$ representations, respectively, of this
$SU(3)$. K\"ahler deformations of the metric $g_{i\bar{j}}$ preserve
the complex index structure.  The metric $g_{i{\jb}}$ itself
transforms as a singlet.  Following \refs{\GranaNY}, we can
use Lefschetz decomposition to 
parametrize K\"ahler deformations according to their $SU(3)$
representations:
\eqn\suthreekd{
	\delta g_{i\bar{j}} = t_{{\bf 1}} g_{i\bar{j}} + (K_{\bf
	8})_{i\jb} \equiv t_{{\bf 1}} g_{i\bar{j}} + t^a_{\bf
	8} (\omega_{{\bf 8}\,a})_{i\jb}.  }
Here $t_{{\bf 1}}$ is a rescaling of the overall volume and transforms
as an $SU(3)$ singlet.  The primitive 2-form $K_{\bf 8}$ transforms in the ${\bf
8}$ of $SU(3)$, and the primitive 2-forms
$\omega_{{\bf8}\,a}=(\omega_{{\bf8}\,a})_{i\jb}dz^i\wedge
d\bar{z}^{\jb}$ for $a=1,\ldots,h^{1,1}-1$ are a basis of the elements
of $H^{(1,1)}(M)$ which each transform in the ${\bf 8}$ of $SU(3)$.

The quantity $\Omega_{ijk}$ is an $SU(3)$ singlet. Therefore
Eqs.~\dvertex\ and \dvertexright\ imply that the auxiliary fields
$y,\hat{\bar{y}}$ transform under the same $SU(3)$ representation as
the K\"ahler deformation in the same supermultiplet.

Let us first consider the auxiliary partners $y_{{\bf 8}}$ of
deformations of type $K_{{\bf 8}}$.  Since $H$ is totally
antisymmetric, a totally holomorphic deformation lies in a singlet of
$SU(3)$.  Therefore $y_{{\bf 8}}$ corresponds to a deformation of the
spin connection only.  It is clear that the deformation
\dvertex\ will break the left-moving $\CN=2$ by breaking the $U(1)_R$
charge.  However, the spin connection couples with the same sign to both the
left- and right-moving fermions. In order that the deformation
preserve the right-moving $\CN=2$, there must be another complex
structure $J_+ = J_{CY} + \delta J_+$ under which the deformation
\dvertex\ is no longer $(3,0)$, and which generates a conserved 
right-moving $U(1)_R$ charge $J_+ = J_{+\mu\nu} \psi_+^{\mu}\psi_+^{\nu}$.

At this point we have a puzzle. If the background described by
$y_{{\bf 8}}\neq 0$ is a manifold, nothing prevents us from defining a
conserved left moving $U(1)_R$ current $\tilde{J}_- =
J_{+\mu\nu}\psi^{\mu}_-\psi^{\nu}_-$ and restoring $\CN=(2,2)$
worldsheet supersymmetry as well as $\CN=2$ spacetime supersymmetry.
The only possible problem is if the global structure of the compactification is
such that $\tilde{J}_-$ is not well defined. This requires transition
functions on the target space of the sigma model which act differently on
the left- and right-moving worldsheet fermions.  This is only possible 
in a locally geometric background if the worldsheet fields
describing different geometric patches are related
by ``stringy" transformations rather than just spacetime diffeomorphisms.
When the background is locally a torus fibration, 
T-duality on the fibers is a classic example of such a transformation.  We will sketch
a more explicit example below.

In summary, for $y_{{\bf 8}} \neq 0$, $\hat{\bar{y}}_{{\bf 8}} = 0$
there are either one or two almost complex structures.  The original
complex structure generates a left-moving $U(1)_R$ that is explicitly
broken and may or may not be globally defined when coupled to either
the left-movers or the right-movers.  There must be a deformed complex
structure which is globally well-defined when coupled to the
right-movers, and is not globally well-defined when coupled to the
left-movers.  The result is a background which is locally described as
having an $SU(3)\times SU(3)$ structure; globally it is not a
manifold, and the left-moving $U(1)_R$ is either broken or not
globally defined. This is consistent with the observation
in \refs{\GranaHR} that compactifications on manifolds with
$SU(3)\times SU(3)$ structure are typically ``nongeometric". Note
again that our methods do not seem to include the totally nongeometric
flux discussed in \refs{\SheltonCF,\SheltonFD}.  We believe that this
is because we are considering a geometric starting point, and studying
small deformations. In this sense, backgrounds that are locally
nongeometric should correspond to some kind of large deformation; it
would be interesting to make this idea more precise.

Next, consider the auxiliary field $y_{{\bf 1}}$ that is the
superpartner of the volume deformation.  In this case, the
corresponding deformation \dvertex\ of the worldsheet can be made up
of both the spin connection $\omega$ and the torsion $H$.  As above,
this preserves $N=(2,1)$ supersymmetry if an almost complex structure
$J_+$ exists which is covariantly conserved with respect to the
torsionful connection $\Gamma_- = \Gamma - \half H$, (where $\Gamma$
is the Levi-Civita connection) and with respect to which the metric is
Hermitian \refs{\HullZY,\GatesNK}.

If $H \neq 0$, the right-moving current $\tilde{J}_- =
J_{+\mu\nu}\psi^{\mu}_- \psi^{\nu}_-$ will not be conserved; this
would require that $\left[\left(\partial + (\Gamma + \half H)\right)
J_+\right]_{\mu\nu\lambda} = 0$, which is incompatible with the
left-moving current being conserved. In this case it is possible that
the background is globally a manifold with $SU(3)\times SU(3)$
structure, without nongeometric features.  Alternatively, $\tilde J_-$
may not be globally well-defined, and we have the same situation as
${\bf y}_{{\bf 8}} \neq 0$ discussed above.

\subsubsec{$\CN=(1,1)$ supersymmetry}

Worldsheet supersymmetry by itself imposes no serious constraints in
this case.  The deformations can lie in the subset of deformations
with $SU(3)$ structure, as long as the complex structure is not
conserved with either connection $\Gamma_{\pm}$.  To be able to tune
$y,\hat{\bar{y}}$ independently, we must be in the class of
compactifications that have a local $SU(3)\times SU(3)$ structure; for
$y_{{\bf 8}}, \hat{\bar{y}}_{{\bf 8}}$ to be independently tuneable,
the backgrounds must presumably be globally nongeometric.

\subsubsec{Auxiliary fields and torsion classes}

We have shown that general values of $y,\hat{\bar{y}}$ correspond to
compactifications with local $SU(3)\times SU(3)$ structure.  The
coupling of the left- and right-moving fermions will be described by
the almost complex structures $J_{\pm}$ defining the two $SU(3)$
structures independently.  Using Eqs.~\scwone, \dvertex,
and \dvertexright, we find that
\eqn\lefttorsion{
\eqalign{
	& \overline W_1^1 + 3 i \overline H_1^1  = y_{{\bf 1}} \cr
	& \overline W_{2,a\bar{b}}^1  = - 8 i \sum_I y^I_{{\bf 8}} \omega^I_{a\bar{b}}
}}
and
\eqn\righttorsion{
\eqalign{
	&W_1^2 + 3 i H_1^2  = \hat{\bar{y}}_{{\bf 1}} \cr
	&W_{2,a\bar{b}}^2  = - 8 i \sum_I \hat{\bar y}^I_{{\bf 8}} \omega^I_{a\bar{b}}
}}
Note that since we are taking our auxiliary fields to be constant
vevs, these equations imply that $W_1$ and $H_1$ are constant. This
seems to imply, by \djdom, that $J \wedge J$ is exact (when all the
other torsion classes are turned off), which is not true for compact Calabi-Yau
backgrounds. In a noncompact model, $J\wedge J$ can be exact (for example,
in flat space). At any rate we only expect to be allowed to turn on a small amount of
torsion and flux in noncompact models.

\subsec{$SU(3)$ structure compactifications}

String compactifications on manifolds with $SU(3)$ structure have been
intensively studied in the last several years. Here we see how these
fit into our framework.  In these cases, $W_k^1 = W_k^2 = W_k$, and
$H_k^1 = H_k^2 = H_k$.  We can see instantly from Eqs.~\righttorsion
and \lefttorsion\ that this means that $y_{{\bf
8}},\hat{\bar{y}}_{{\bf 8}}$ are complex conjugates.

Because one can deform the metric and the $H$-flux independently, we
may still separately dial $y_{{\bf 1}}, \hat{\bar{y}}_{\bf 1}$.  One
might be tempted to conclude that since we can set $y_{{\bf 1}} \neq
0$, $\hat{\bar{y}}_{{\bf 1}} \neq 0$, there is a class of $SU(3)$
structure compactifications with $W_1 \neq 0$ which is compatible with
spacetime supersymmetry.  However, we will see in \S4\ that the
auxiliary field for the universal (dilaton) hypermultiplet is a
different linear combination of $W_1$ and $H_1$.  Therefore, if
$y_{{\bf 1}} \neq 0$ and $\hat{\bar{y}}_{{\bf 1}} = 0$, the auxiliary
field $\hat{\bar{y}}_{\phi}$ in the universal hypermultiplet will
still be nonzero.  This matches the known fact that half-flat
manifolds in type IIB are incompatible with spacetime supersymmetry
(see \refs{\GranaJC}\ and references therein).

\subsec{Nongeometric compactifications}

We have shown that typical compactifications with $y_{{\bf 8}} \neq 0$
should have ``nongeometric" features, in which the transition
functions on different coordinate patches act chirally on the
fermions.  Our argument was based on the requirement that despite the
existence of a well-defined left-moving $U(1)_R$ current constructed
from an almost complex structure, there should be no well-defined
right-moving $U(1)_R$ current constructed from the same almost complex
structure.  We will sketch a scenario in which this can occur, to make
our reasoning clearer.

Consider a compactification that is locally a complex manifold with a
Lagrangian $T^3$ fibration.  This fibration will have monodromies as
one encircles singular loci of the fibration on the base
$B$ \refs{\TomasielloBP,\StromingerIT,\MorrisonBT}. For a purely
geometric fibration, such as a Calabi-Yau \refs{\StromingerIT}\ or a
manifold with ``geometric
flux" \refs{\TomasielloBP,\SheltonFD,\KachruHE,\GranaKF}, the
monodromy will lie in the group $GL(3,\QZ)$ of discrete
diffeomorphisms.

More generally, this monodromy can lie in the full duality group of
$T^3$.  We will restrict ourselves (arbitrarily) to the perturbative
duality group $O(3,3; \QZ)$.  Such backgrounds are often known as
``T-folds"; various examples have been discussed in
\refs{\DabholkarSY\DabholkarVE\HullHK\HellermanAX\HullIN
\LawrenceMA\FlournoyVN-\FlournoyXE}

Imagine that the monodromy about some loop is a non-trivial T-duality
acting on two 1-cycles of the torus, much as in the example discussed
in \refs{\KachruHE}.  Let the coordinates $y^i$ be the coordinates on
the $T^3$ and $x^i$ be coordinates on the base $\CB_3$, such that $z^i
= x^i + i y^i$, for $i=1,2,3$. If the T-duality in the monodromy
discussed above acts on $y^{2,3}$, then it will act on the fermions as
\eqn\tdfermionact{
	\psi_{\pm}^{y^{2,3}} \to \pm \psi_{\pm}^{y^{2,3}}
}
and all other worldsheet fermions lying along the compactification
directions will not transform.  If we write the fermions in complex
coordinates, then the monodromy will act as:
\eqn\monodcomplex{
\eqalign{
	\psi_+^{z^i} & \to \psi_+^{z^i} \cr
	\psi_{-}^{z^1} & \to \psi_{-}^{z^1} \cr
	\psi_-^{z^{2,3}} & \to \psi_-^{\bar{z}^{2,3}}
}}

Now, suppose that
\eqn\leftuone{
	J_+ = \sum_{i = 1}^3 \psi_+^{z^i} \psi_+^{\bar{z}^i}
}
is a globally well defined operator---in other words, it will be
preserved under monodromy transformations arising from loops in
$\CB_3$ about the singular loci of the fibration. (The operator is
certainly invariant under the monodromy action above). It may seem
natural to define a left-moving $U(1)_R$ current
\eqn\rightuone{
	J_- = \sum_{i=1}^3 \psi_-^{z^i} \psi_-^{\bar{z}^i}
}
but this is not globally well-defined, as the monodromy action on
$J_-$ is
\eqn\monoduone{
	J_- = \psi_-^{z^1}\psi_-^{\bar{z}^1} +  \psi_-^{z^2}\psi_-^{\bar{z}^2} +  
	\psi_-^{z^3}\psi_-^{\bar{z}^3} \to 
	 \psi_-^{z^1}\psi_-^{\bar{z}^1} -  \psi_-^{z^2}\psi_-^{\bar{z}^2} -  \psi_-^{z^3}\psi_-^{\bar{z}^3}
}

In general, only a monodromy action which includes a T-duality
transformation will act chirally on the fermions in this way.  A more
global analysis is required to see if a different left-moving
$U(1)_R$ charge is globally defined, and whether it is conserved; if
no such conserved current exists, then one can have an $\CN=(1,2)$
compactification without $H$-flux, as seems to appear when only the
$y_{{\bf 8}}$ auxiliary fields are turned on.

\subsec{Mirror symmetry}

One goal of this paper is to understand mirror symmetry for
compactifications with H-flux.  Here we argue that our results confirm
previous
statements \refs{\GranaSN,\GranaNY\GranaHR,\SheltonCF,\SheltonFD}\
that the mirrors of such compactifications are compactifications with
local $SU(3)\times SU(3)$ structure, which are often T-folds.

The first argument arises from the four-dimensional effective action.
Consider a type IIA compactification with nonvanishing auxiliary
fields for the hypermultiplets, which are complex structure
deformations.  In \refs{\LawrenceZK,\LawrenceKJ}, these auxiliary
fields were shown to be combinations of $H$-flux of type $H_{3,4}$ and
torsion of type $W_{3,4}$, according to the definitions \djdom\
and \hexp.  Mirror symmetry should leave the variables of the
four-dimensional effective action invariant.  Thus, we expect that the
mirror of such combinations of flux and torsion to be auxiliary fields
of type $y_{1},y_{{\bar 8}}$, which are related to the K\"{a}hler moduli.

The essence of this argument is that if mirror symmetry holds for
Calabi-Yau backgrounds, it holds for deformations of Calabi-Yau
backgrounds, as long as one understands the mirror map acting on
deformations of the theory.  Of course this is dangerous; we are
assuming that compactifications of the type we care about can be
considered as connected in field space to an $\CN=2$ compactification,
and that mirror symmetry remains valid for off-shell deformations.
Since we expect mirror symmetry to hold for all correlation
functions (which can be used to define an effective potential, at least in
a power series about a given point in field space) the
assumption is not obviously wrong.

The second argument arises from the worldsheet point of view.  From
this standpoint the reader might sensibly object that the cases we
have in mind have at most $\CN=1$ spacetime supersymmetry and therefore at most
$\CN=(2,1)$ worldsheet supersymmetry. The conformal field theory explanation of
mirror symmetry uses the structure of the $\CN=(2,2)$ superconformal
algebra: namely, one simply reverses the sign of the right-moving
$U(1)_R$ current relative to the left moving $U(1)_R$ current.

However, we do not believe that this poses an obstacle to defining
mirror symmetry from the worldsheet point of view, at least if the IIA
compactification is a manifold with $H$-flux. Such type IIA
compactifications with $y \neq 0$ still have a right-moving $U(1)_R$
current $\tilde{J}_+$ which exists but is not conserved: take the
almost complex structure $J_{-\,\mu\nu}$ used to define the
left-moving $U(1)_R$ current, and write $\tilde{J}_+ =
J_{-\,\mu\nu} \psi^{\mu}_+\psi^{\nu}_+$.  It still makes sense to
reverse the sign of this non-conserved $U(1)_R$.  In the IIB mirror,
while the local almost complex structure defining the left moving
$U(1)_R$ cannot be used to construct a globally well defined
right-moving $U(1)_R$ current, there would be a different local almost
complex structure which leads to a globally defined but {\it
non\/}conserved right-moving $U(1)_R$ current.

\newsec{Ten-dimensional supergravity calculation of auxiliary fields}

\def\CV{{\cal V}}

%\HassanBV
\lref\HassanBV{
  S.~F.~Hassan,
  ``T-duality, space-time spinors and R-R fields in curved backgrounds,''
  Nucl.\ Phys.\  B {\bf 568}, 145 (2000)
  [arXiv:hep-th/9907152].
  %%CITATION = NUPHA,B568,145;%%
}

In this section, we will calculate the auxiliary fields using ten
dimensional supergravity, by studying the supersymmetry variations of the
fermions in the hypermultiplets directly. The four-dimensional supersymmetry
transformations are \mwdecompose, with $\eta^N$ the nowhere-vanishing
spinors defining the $SU(3)\times SU(3)$ structure, and $\eta_1 = \eta_2$ for
Calabi-Yau backgrounds. For the K\"ahler moduli, this
calculation should be essentially identical to the calculations in
Sec.~3---after all, the formula \auxfieldope\ essentially implements
two spacetime supersymmetry transformation on the worldsheet vertex
operator for the scalar in the multiplet
\refs{\AtickGY,\DineGJ}.  One difference will be that in this section
we also consider auxiliary fields for the universal hypermultiplet,
with a superfield expansion as given in \twohyper.  It would be
worthwhile to do the same calculation for the auxiliary fields that
arise in the presentation of \refs{\BerkovitsCB}, or for other
off-shell hypermultiplets. Additionally, now that we are 
no longer confined to a worldsheet description, one could include RR fluxes (as in \GranaNY). 
We expect these fields will show up as additional contributions to the auxiliary fields; an 
argument that the RR 0-form appears in such a manner for vector multiplets was given in 
\refs{\LawrenceZK,\LawrenceKJ}. They will also appear in other off-shell representations of the hypermultiplets. We leave this exercise for future work.

We now proceed to calculate $y_{{\bf 1}}$, $y_{{\bf 8}}$, and the
auxiliary field $y_{{\phi}}$ for the universal hypermultiplet, in
turn.

\subsec{Auxiliary field for the volume deformation $y_{{\bf 1}}$}

For a general K\"ahler deformation \suthreekd\ of the Calabi-Yau
metric away from a fixed metric $g_{i\bar{j}}$, the rescaling $t_{{\bf
1}}$ of the volume can be picked out by contracting $\delta g$ with
$g$ ,
\eqn\tracepart{
	t_{{\bf 1}} = \frac{1}{3} g^{i\bar{j}} \delta g_{i\bar{j}}\ .
}
Since the trace will select out a variation that is proportional to
$\delta g$, it will be independent of the internal coordinates.

We will work in ten-dimensional string frame, to more easily match the
worldsheet calculation of the previous section.  The supersymmetry variation of
$\delta g$ implies that the fermionic superpartner is:

\eqn\fermionpartnervolume{
	\chi^N_{{\bf 1}\pm}  = \frac{i}{3} (\eta^{N}_{\pm})^\dagger \gamma^{\bar i} \Psi^N_{\bar i}
}
where the $\pm$ subscripts for $\chi$ denote four-dimensional chirality,
$\Psi^N_m$ are the two ten-dimensional gravitinos polarized along $M$,
and we have used $\delta_{\eps^N} (\delta g_{mn}) = i \bar \eps^N
(\Gamma_m \Psi_{n}^N + \Gamma_n \Psi_m^N )$.  The bosonic part of the supersymmetry
variation of $\chi_{{\bf 1}}$ will come entirely from the supersymmetry
variation of $\Psi_m$ in \fermionpartnervolume. These are given in
\refs{\PolchinskiRR,\HassanBV}.\foot{Note that the subscripts $\pm$ in
Appendix~B of Ref.~\HassanBV\ label the two supersymmetries
of \hbox{type IIB} supergravity.}  Let $y^{N=1}_{{\bf 1}} = y_{{\bf
1}}$ and $y^{N=2}_{{\bf 1}} = \hat{\bar y}_{{\bf 1}}$.  The supersymmetry
variation of $\psi_{{\bf 1},+}$ with respect to an infinitesimal
parameter $\zeta$ is:
\eqn\volfermvar{
\eqalign{
	\delta \chi^{N}_{{\bf 1}+} & = 
	 \zeta_{N+} \otimes y^N_{{\bf 1}}  + 
	\ldots\cr
	& = \frac{i}{3} \zeta_{N+} \otimes (\eta^{N}_+)^\dagger \gamma^{m} 
	\left( D_{m} + (-1)^N \frac{1}{8} H_{mBC} \gamma^{BC}\right) \eta^N_{-} + \cdots
}}
where $D_m \eta = \p_m \eta
+ \frac{1}{4} \omega_{mAB} \gamma^{AB} \eta$ and capital letters
$A,B,\ldots$ run over all six internal indices.

Now, using \spinorscovder, we find that
\eqn\auxvolstepone{
	y_{{\bf 1}} = \frac{i}{3} \left [ iq_{mn}(\eta^{\dagger}_+ \gamma^{m}   \gamma^n \eta_+)
	- \frac{1}{8} \eta^{\dagger}_+ \gamma^{m} \gamma^{BC} H_{m BC} \eta_- \right ]
}
The specific expressions for $q$ are given
in \refs{\GranaBG,\FidanzaZI}.  ($q_m, \tilde{q}_m$ do not contribute
because they multiply the wrong chirality).  Using \suthreeacs, the
first term on the left hand side of \auxvolstepone\ will project on
the $SU(3)$ singlet part of $q_{mn}$. The result, combined
with \suthreetopform, confirms \lefttorsion, \righttorsion\ precisely
for $y_{{\bf 1}}, \hat{\bar y}_{{\bf 1}}$, up to an overall prefactor.

\subsec{Auxiliary fields $y_{{\bf 8}}$ for remaining K\"ahler moduli}

For a general deformation of the metric $\Delta g_{mn}$ polarized
along the internal directions, the fermion partner under the supersymmetry
deformation can be computed using the ten-dimensional supersymmetry algebra:
\eqn\genpartner{
	\chi^N_{(mn),\pm} = (\eta^{N}_{\pm})^\dagger (\gamma_m \Psi^N_n + \gamma_n \Psi^N_m )\ .
}
Note that these are not yet four-dimensional fields in the usual
sense; $\Delta g$ and $\psi$ depend on the coordinates of $M$.  For a
particular four-dimensional scalar arising from reducing the metric on
a particular internal wavefunction, we would reduce the fermion as
well.

To find the auxiliary field $y$ for such a deformation, we compute
$\delta\chi^1_+$ and find that:
\eqn\genaux{
	y^1_{(mn)} = \left(i \delta_{(m}{}^p - J_{(m}{}^p\right)q_{n)p} - i \Omega_{(m}{}^{pq}H_{n)pq}
}
Again, $y^1_{(mn)}$ as defined is not yet a four-dimensional field in
the usual sense.

Now, if we are interested in the auxiliary fields for $\Delta
g_{i\bar{j}} = t^I_{{\bf 8}} \omega^I_{i\bar{j}}$, with $\omega^I$ a
basis of primitive harmonic $(1,1)$ forms, we should expand $y^1 = y$
in the same basis.  Following \refs{\GranaBG,\FidanzaZI}, $H$ has
terms transforming in the $1$, ${\b 3}$, ${\bf 6}$, and conjugates,
and will not contribute.  The only term in $q_{mn}$ transforming in
the ${\bf 8}$ is proportional to $W_2$, and the result
confirms \lefttorsion,\righttorsion\ for $y_{{\bf 8}}$.

\subsec{Auxiliary field $y_{\phi}$ for the universal hypermultiplet}

The real part of the scalar field in the universal hypermultiplet is
the four-dimensional dilaton $\phi_4 = \phi_{10}
- \half \ln \CV$. Since $\Delta \CV = 3 \Delta t_{{\bf 1}} \CV$, we
find that the four-dimensional dilatino is, in string frame
\eqn\fourddilatino{
	\lambda^{\phi_4,N}_{\pm} = (\eta^{N}_{\pm})^\dagger \lambda^{\phi_{10},N}
		- \frac{3}{2} \chi^N_{{\bf 1},\pm}
}
Using the variations given in \refs{\PolchinskiRR,\HassanBV}, we find that
\eqn\auxfielddil{
\eqalign{
	y_{\phi_4} & = - \frac{3i}{4} \left(\bar{W}^1_1 + i \bar{H}^1_1\right) \cr
	\hat{\bar{y}}_{\phi_4} & = - \frac{3i}{4} \left(W^2_1 - i H^2_1\right)
}}

As an application of this result, recall that in \S3.4, we commented
that even though one can set $y_{{\bf 1}} = 0$ in the set of
compactifications with $SU(3)$ structure (for which $H^1_1 = H^2_1 =
H_1$, $W^1_1 = W^2_1 = W_1$), we expect that supersymmetry is broken
if $W_1,H_1 \neq 0$. This is because $y_{{\bf 1}}$ and $y_{\phi_4}$
are independent linear combinations of $\bar{W}_1,\bar{H}_1$, and can
both vanish only if $W_1,H_1 = 0$.

This is very close to the results in \refs{\GranaNQ,\CamaraKU}, which
find that the F-term for the dilaton is a combination of the $(3,0)$ components
of the NS-NS and R-R 3-form field strengths.  We find a purely NS-NS deformation
because we are studying auxiliary fields which break a different $\CN=1$ subgroup of
the $\CN=2$ supersymmetry.

\newsec{Four-dimensional Supergravity calculation of auxiliary fields}

\def\b{\beta}\def\e{\epsilon}\def\o{\omega}
\def\w{\wedge}
\def\CN{{\cal N}}

\def\IIB{{\rm IIB}}
\def\Vol{{\rm Vol}}

In this section we check our results in \S3.4\ against the superpotential
for $SU(3)$ structure compactifications proposed by \refs{\GranaNY}.
Closely related and complementary results were obtained in \refs{\CamaraCZ}.

\subsec{Superpotential for $SU(3)$ structure compactifications}

Gra\~na {\it et. al.}\ \refs{\GranaNY}\ computed the $\CN=2$ Killing
prepotential and the superpotential for an arbitrary $\CN=1$
subalgebra of the $\CN=2$ symmetry of the effective action, for
compactifications with $SU(3)\times SU(3)$ structure. We will specialize to
the case of $SU(3)$ structure with one or more of $\CW_{1,2}$ and
$H_1 \neq 0$, and focus on the $\CN=1$ subalgebras generated by the
left- and right-movers on the worldsheet.

As we have pointed out in the previous two sections, $SU(3)$ structure
compactifications of this type break supersymmetry completely.  In
order to check our previous formulae, we will choose one of the
$\CN=1$ subalgebras and use the results of \refs{\GranaNY}, who
extract the corresponding superpotential from the prepotentials.  For
example, we can choose a superpotential $W$ for the $\CN=1$ supersymmetry
corresponding to the superspace directions $\theta$; this implies a
nonvanishing expectation value of $\hat{\bar y}$.  The auxiliary field
$y$ of the $\CN=2$ hypermultiplet becomes the auxiliary field for the
$\CN=1$ chiral multiplet
\eqn\nonechiral{
	w^a + \theta \chi^a + \theta^2 y^a
}
We can determine $y$ by computing the K\"ahler covariant derivative of
$\bar W$:
\eqn\FfromW{F^A = -e^{K/2 m_{p,4}^2}K^{a\bar b}D_{\bar b}\bar W\ ,}
where $K$ is the K\"ahler potential, and $K^{a\bar{b}}$ is the
inverse of $K_{a\bar{b}} = \p_a \p_{\bar{b}} K$, and $m_{p,4}$ is the four-dimensional Planck mass. The K\"ahler
covariant derivative is: 
\eqn\kcovder{
	D_a W = \p_a W + \frac{1}{m_{p,4}^2}W \p_a K \ .
}

The results of \refs{\GranaNY}\ can be summarized as follows. Let us
denote K\"ahler and complex structure moduli by $t$ and $u$,
respectively.  Define
\eqn\Wplusminus{W(t,u) = i m_{p,4}^3 \int(B+iJ)\w d\Omega
\quad\hbox{and}\quad
\tilde W(\bar t,u) = -i m_{p,4}^3 \int(B-iJ)\w d\Omega.}
where we assume that all geometric quantities in the integral are
given in string units, and the dimensionful factor in front
is the correct one for the four-dimensioinal effective action (\cf\ the
appendix of \refs{\DenefMM}).\foot{In our noncompact case, one might worry about the convergence of \Wplusminus. To avoid such problems, we should think cutting off the large-volume part of our Calabi-Yau and gluing it into a compact space, taking appropriate care with boundary conditions. We leave such questions to future work.}
Specializing to the case of $SU(3)$ structure, the results of
\GranaNY\ (cf.~Eqs.~(2.148) and (2.149)) imply that the $\CN=1$
superpotentials for the $\CN=1$ subalgebras generated by left- and
right-moving supersymmetries, respectively, are:
\eqn\WIIAIIB{
W_\IIB(t,u) = W(t,u),
\quad
\hat{W}_\IIB(\bar t,u) = \tilde W(\bar t, u)\ .}
These can be checked by matching the coefficients of the H-flux in the
gravitino and dilatino variations, between variations of the string frame fields
in \refs{\PolchinskiRR,\HassanBV} on the one hand and the variations of the
Einstein frame fields in \refs{\GranaNY}.

For pertubations away from a Calabi-Yau geometry, these formulae
deserve a word of interpretation.  We wish to split variations of $B,
J$ into:
\eqn\splitvar{
\eqalign{
	B & = B_{CY} + B_t\cr
	J & = J_{CY} + J_t
}}
Here $J_{CY}, B_{CY}$ correspond to the K\"ahler form and \NSNS\ 2-form
for the underlying Calabi-Yau compactification -- in particular we
take them to carry all of the dependence on the K\"ahler moduli, and
$d B_{CY} = d J_{CY} = 0$.  $B_t, J_t$ are not closed -- $dB_t = H$,
and $dJ_t$ is a proportional to the intrinsic torsion. Similarly, we
will write $\Omega = \Omega_{CY} + \Omega_t$, where $\Omega_{CY}$ is
closed and $d\Omega_t$ is proportional to various intrinsic torsion
classes.  In particular, if we assume the intrinsic torsion and $H$
are small, then we can write
\eqn\torsionvariation{
\eqalign{
	dJ_t & = \frac{3i}{4} \left(W_1 \bar \Omega_{CY} - \bar{W}_1 \Omega_{CY}\right)
		+ W_4 \wedge J_{CY} + W_3\cr
	dB_t & = H_t = \frac{3i}{4} \left(H_1 \bar \Omega_{CY} - \bar{H}_1 \Omega_{CY}\right)
		+ H_4 \wedge J_{CY} + H_3\cr
	d\Omega_t & = W_1 J_{CY}^2 + W_2\w J_{CY} + \bar{W}_5 \w \Omega_{CY}.
}}
Note that this decomposition into a background and deformation is only valid for a noncompact Calabi-Yau, where we are allowed to locally turn on small amounts of torsion. In general, such a decomposition would not be sensible. 

Let us consider the $\CN=1$ algebra generated by the left-movers, with
superpotential $W$. The $\CN=1$ K\"ahler potential is the sum of terms
$K_1$, $K_2$ and $K_3$ for the K\"ahler moduli, complex structure
moduli and dilaton, respectively. The K\"ahler potential for K\"ahler
moduli is:
\eqn\Kone{e^{-K_1/m_{p,4}^2} = {4\over3}\int J\w J\w J = 8 V,}
where $V$ is the volume of $M$ in string units. The K\"ahler potential
for complex structure moduli is
\eqn\Ktwo{e^{-K_2/m_{p,4}^2} = i\int\Omega\w\bar\Omega = 8V.}
where $\Omega$ is the canonically normalized (3,0)-form ($\Omega_\eta$
in the notation of Ref.~\GranaNY).\foot{Canonically, ${1\over6}J^3 =
{i\over8}\Omega\w\overline \Omega = \Vol_6$, where $\Vol_6$ is the
volume form.  In the framework of Ref.~\GranaNY, the K\"ahler and
complex structure pure spinors are $\Omega_+ = ce^{-B-iJ}$ and
$\Omega_- = n\Omega$.}  Finally, the K\"ahler potential for the
four-dimensional dilaton is:
\eqn\Kthree{e^{-K_3/m_{p,4}^2} = e^{-2\phi_4} = -{i\over2}(\tau-\bar\tau),}
where $\tau = a + ie^{-2\phi_4}$'; $a$ is the model-independent \NSNS\
axion dual to the \NSNS\ 2-form in four dimensions; and $\phi_4$ is the
four-dimensional dilaton.

\subsec{Overall volume modulus: {\bf 1} of $SU(3)$}

We consider the metric deformation $t_{{\bf 1}}$ in the ${\bf 1}$ of
$SU(3)$: that is, $\delta J = t_{{\bf 1}} J$. The complex scalar in
the chiral multiplet can be written as $w = b + i t_{{\bf 1}}$, where
$B = b J$.  If we write $V = V(w - \bar{w})$, then
\eqn\volumescaleder{
	\p_w V|_{w = 0} = - \p_{\bar{w}} V|_{w = 0} = - \frac{3i}{2} V\ .
}
The relevant derivatives of the K\"ahler potential are:
\eqn\kahlerderivs{
\eqalign{
K_{1, \bar{w}} & = - m_{p,4}^2 \p_{\bar{w}} \ln V = -\frac{3i}{2} m_{p,4}^2\cr
K_{1, w\bar w} & = - m_{p,4}^2 \p_w \p_{\bar{w}} \ln V = \frac{9}{4} m_{p,4}^2
}}
We wish to calculate $y_{{\bf 1}}$, using \FfromW\ with $W =
W_{IIB}(t,u)$. The K\"ahler covariant derivative of $\bar{W}$ is:
\eqn\superpotderivs{
\eqalign{
D_{\bar {w}} \bar W &= - i m_{p,4}^3 \partial_{\bar{w}} \int(B_{CY}-iJ_{CY})\w d\bar{\Omega}_t
+ \frac{3}{2} m_{p,4}^3 \int(H_t - idJ_t)\w\bar{\Omega}_{CY}\cr
& = -i m_{p,4}^3 \int J_{CY} \w d\bar{\Omega}_t 
+\frac{3}{2} m_{p,4}^3 \int(H_t - idJ_t)\w\bar{\Omega}_{CY}\cr
& = - i \bar{W}_1 \int J^3 + \frac{3}{2} \left(\frac{-3i}{4}\right) \left(\bar{H}_1 - \bar{W}_1\right)
	\int \Omega_{CY}\wedge \bar{\Omega}_{CY}\cr
& = a V \left(\bar{W}_1 + 3 i \bar{H}_1\right) 
}}
where we have used $J_{CY}^3 = \frac{3i}{4} \Omega_{CY}\w \bar{\Omega}_{CY}$
(\cf \refs{\GranaNY}), and where $a$ is a numerical constant.  The resulting
auxiliary field is:
\eqn\Fw{y^w \equiv F^w \propto e^{\phi_4} m_{p,4} 
\left( \bar{W}_1 + 3i \bar H_1\right) = m_{s} \left(\bar W_1 + 3 i \bar H_1\right)\ .}
This matches \lefttorsion.  Note the explicit
factor of the string mass $m_{s}$. The auxiliary partner $F$ of a
dimensionless scalar will have mass dimension 1.  The discussion
of \S3,\S4 was entirely in string frame; all of the lengths were
measured in string frame. The dimensions of superpartners arise from
explicit powers of $m_{s}$ that appear in the spacetime superalgebra
in string frame.

\subsec{Primitive K\"ahler moduli: {\bf 8} of $SU(3)$}

We now consider arbitrary K\"ahler moduli $w^a$ in the ${\bf 8}$ of
$SU(3)$, defined by $B + i J = \sum_a w^a \omega^a$, where $\omega^a$
is a basis of $H^{1,1}$. The derivatives of the K\"ahler potential
are:
\eqn\kderivgenk{
K_a = -{im_{p,4}^2 \over4V}\int\o_a\w J^2,\quad
K_{a\bar b} = -{m_{p,4}^2 \over4V}\int\o_a\w\o_b\w J
+ {m_{p,4}^2 \over 16V^2}\left (\int\o_a\w J^2 \right ) \left (\int\o_b\w J^2\right )\ ,}
and the derivative of the superpotential is:
\eqn\genKsuperder{
	\partial_a W = im_{p,4}^3 \int\o_a\w d\Omega 
= im_{p,4}^3 \int\o_a\w (W_2\w J + W_1 J^2)\ .
}

For K\"ahler moduli $w^a$ in the {\bf 8} of $SU(3)$, the corresponding
$(1,1)$ forms $\o_a$ are primitive. so that $\o_a\w J^2 = 0$.  In this
case,
\eqn\primitivekahlerder{
K_a = 0,\quad K_{a\bar b} = -{m_{p,4}^2 \over4V}\int\o_a\w\o_b\w J
\quad\hbox{and}\quad
\partial_a W = im_{p,4}^3 \int\o_a\w W_2\w J\ .}
Writing $\bar W_2 = {\bar W_2}^b\o_b$, 
\eqn\primitivecovder{
	D_{\bar a}\bar W = 4iV K_{{\bar a}b} m_{p,4} \bar W_2^b\ .}
so that
\eqn\Fwa{y^a = c_{{\bf 8}}
e^{\phi_4}m_{p,4} \bar W_2^a = c_{{\bf 8}} m_s \bar W_2^a\ .}
where $c_{{\bf 8}}$ is some complex numerical coefficient.  Similarly,
\eqn\Fwa{{\hat{\bar y}}^a =
c_{{\bf 8}}^* m_s W_2^a\ .}
This confirms the results given in \S3 and \S4.

\subsec{Universal hypermultiplet}

Next, let us describe the auxiliary fields of the universal
hypermultiplet.  Since Eq.~\Wplusminus\ is independent of the dilaton,
we have $D_\tau W = K_\tau W$ proportional to the superpotential.  In
type IIB, we find that
\eqn\DilAuxFields{\eqalign{y^\tau &= c_{\phi,1} e^{-\phi_4}m_{p,4}
\left( \bar H_1 - i\bar W_1\right) \ ,\cr
\hat{\bar y}^\tau & = c_{\phi}^* e^{-\phi_4}m_{p,4} 
\left( H_1 - i W_1\right) \ .}}
where $c_{\phi}$ is a complex numerical constant.
Now to compare this to previous sections, we need to transform
$y^{\tau}$ to $y^{\phi}$, using $\phi = -\half \ln[(\tau
- \bar{\tau})/2i]$. We find that
\eqn\dilauxtwo{\eqalign{
	y^\tau &= \tilde{c}_{\phi} m_s \left(
\bar H_1 - i\bar W_1\right) \ ,\cr
\hat{\bar y}^\tau & = \tilde{c}_{\phi}^* m_s 
\left( H_1 - i W_1\right) \ .,}}
where $\tilde{c}_{\phi}$ is some complex numerical coefficient. This matches the
result in \S4.3.

\subsec{Complex structure moduli: {\bf 6} of $SU(3)$}

Finally, in order to tie the language of $SU(3)$ structures to
previous work \refs{\LawrenceZK,\LawrenceKJ}, we will also compute the
\NSNS\ auxiliary fields for the type IIB complex structure moduli.

Refs.~\refs{\LawrenceZK,\LawrenceKJ}\ showed that the auxiliary fields
$D_{++}$ corresponded to $H$ and $dJ$ lying in $H^{(2,1)}$.  From the
complex structure moduli $u^A$ in the {\bf 6} of $SU(3)$, we define a
basis of primitive (2,1) forms
\eqn\twooneforms{\chi_A = D_A\Omega = (\partial_A+K_A)\Omega.}
$H$ and $dJ$ can be expanded in this basis, as we will do.

In terms of the $\chi_A$ and their complex conjugates, the metric
$K_{\bar A B}$ in the complex structure moduli space is
\eqn\WPmetric{K_{\bar A B} =
\bar\partial_{\bar A}\partial_{B} K
= -{\int\chi_B\w\bar\chi_{\bar A}\over \int\Omega\w\bar\Omega}
= -{i\over8V}\int\chi_B\w\bar\chi_{\bar A}.}
Writing the real 3-forms $H_3$ and $W_3$ as
\eqn\HWthree{H_3 = H_3^A\chi_A + \bar H_3^{\bar A}\bar\chi_{\bar A}
\quad\hbox{and}\quad
W_3 = W_3^A\chi_A + \bar W_3^{\bar A}\bar\chi_{\bar A},}
we can use \twooneforms,\FfromW, and \Wplusminus, to show that:
\eqn\Fcpx{D_{++}^A = c_{{\bf 6}} m_s \left( \bar{H}_3^A-i\bar{W}_3^A\right) 
\quad\hbox{and}\quad
\hat D_{--}^A = c_{{\bf 6}} m_s \left( \bar{H}_3^A+i\bar{W}_3^A\right) \ .}
for the \NSNS\ auxiliary fields in the vector multiplets, where $c_{\bf{6}}$ is a complex numerical
constant. This is in precise agreement with the
results in \refs{\LawrenceZK,\LawrenceKJ}.

\newsec{Worldsheet instanton corrections}

\subsec{General remarks}

Supergravity arguments have indicated that the mirror of
compactifications with $W_3,H_3 \neq 0$ involves intrinsic torsion
classes in a locally $SU(3)\times SU(3)$ structure background
\refs{\GranaNY,\GranaSN,\GranaHR}.  While this identification is
surely correct, we expect that supergravity will be a poor
approximation for such compactifications.\foot{The astute reader will
sensibly complain that this has not stopped us from doing supergravity
calculations either.  Again, for local models we have some hope of
being on good footing; beyond that, we should start from the
worldsheet discussion in \S3.}

One reason is that, as we have argued in \S3, nongeometric fluxes are
generic features of the compactification. The second reason arises
from contemplating mirror symmetry in its traditional setting, type II
compactifications on Calabi-Yau backgrounds.  For compact models
mirror symmetry only makes sense when worldsheet instantons are
included---indeed, the ability to compute instanton effects is to a
large extent what made mirror symmetry so exciting in the first place.

More precisely, let us consider type IIA string theory with
expectation values for auxiliary fields in the hypermultiplets.  These
will be described by \NSNS\ flux and torsion of type $H_3$ and $W_3$
which, as mentioned in \S5.5, leads to a superpotential for complex
structure moduli.  For a compact model the minimum of the
corresponding potential is generically deep in the interior of complex
structure moduli space.  If mirror symmetry is at all valid, the
mirror should have volumes of order the string scale, for which
nonperturbative worldsheet physics should become important.

Furthermore, recall that in local (noncompact) models, the
superpotential for complex structure moduli arises as a term breaking
$\CN=2$ supersymmetry to $\CN=1$ supersymmetry.  For vector multiplet
moduli, one expands the prepotential to first order in the
nonvanishing auxiliary field, and integrates out the superspace
directions for the broken supersymmetry, to obtain the superpotential
\refs{\LawrenceZK,\LawrenceKJ,\VafaWI,\TaylorII}.  A similar
calculation should hold for the hypermultiplet moduli.  While we leave
this project for future work, we note that the $\CN=2$ action for the
K\"ahler moduli will receive worldsheet instanton corrections, and so
we expect the superpotential to receive such corrections as well.
Indeed, in Ref.~\GurrieriIW, Gurrieri and Micu attempted to match the
bosonic four-dimensional effective action for type IIB string theory on a half-flat
manifold to a type IIA compactification with \NSNS\ 3-form flux.
Close inspection of this paper reveals that in order for the actions
match in detail, coefficients of various terms in the type IIB
effective action must include terms from worldsheet instanton
corrections.

\subsec{One-instanton contribution to the superpotential}

We wish to show that when one can perturb a Calabi-Yau background to
an $SU(3)$-structure background with intrinsic torsion of type
$W_{1,2}$, worldsheet instantons contribute to the superpotential (for
the $\CN=1$ spacetime supersymmetry generated by either the left- or
right-moving worldsheet sector), with the one-instanton contribution
entering precisely at first order in $W_{1,2}$.

We begin by reviewing the argument in \refs{\DineZY,\DineBQ}, that for
string backgrounds with $\CN=(2,2)$ worldsheet supersymmetry, the
K\"ahler moduli do not get instanton-generated superpotentials. We
will adopt the specific arguments in \refs\KachruIH\ to our present
purposes.

Assume that at least $\CN=1$ spacetime supersymmetry is preserved, and
arises from the right-moving $\CN=2$ worldsheet algebra (this could be
secretly $\CN=2$ supersymmetric, or the $U(1)_R$ charges of the
right-moving vertex operators could fail to satisfy the correct
quantization conditions \refs{\BanksCY,\BanksYZ}.) The supersymmmetry
transformations generate superspace translations along $\hat\theta$,
and we will be studying antichiral superfields with respect to this
supersymmetry, to match the expansion in \twohyper.  If the K\"ahler
moduli $t^a$ are the real parts of scalar components $\phi^a$ of term
of $N=1$ superfields
\eqn\expand{
	\Phi^a = \phi^a + \hat{\bar{\theta}} \psi^a + \hat{\bar{\theta}}^2 \hat{\bar{y}}^a
}
then a superpotential of the form $W = \phi^a \Phi^b \Phi^b$ leads to
terms of the form $\phi^a \phi^b \hat{\bar y}^c$ in the low-energy
action, where $\hat{\bar y}^i$ is the auxiliary field.  This term
exists if the worldsheet correlator
\eqn\threepointcorr{
	A = \langle V_{\phi^a}^{(-1,-1)} V_{\phi^b}^{(-1,-1)} V_{\hat{\bar{y}}^c}^{(0,0)} \rangle
}
where the superscripts refer to the superconformal ghost charge (or,
the picture of the vertex operator), and the subscripts to the
corresponding spacetime fields. The zero-momentum $(-1,-1)$ picture
scalar vertex operators are given in \wvertex,
and $V_{\hat{\bar{y}}}$ is shown in \dvertexright.

Therefore, there are six total worldsheet fermions appearing
in \threepointcorr.  However, in the one-instanton sector of two-dimensional
$\CN=(2,2)$ theories, there are eight fermion zero
modes \refs{\DineZY,\DineBQ}: four left-moving fermions, two with
holomorphic spacetime indices and two with antiholomorphic target
space indices; and four right-moving fermions, again, two with
holomorphic spacetime indices and two with antiholomorphic spacetime
indices.

Now imagine that we can slightly deform the Calabi-Yau metric such
that the new metric has torsion of the type $W_{1,2}$.  In particular
imagine giving an expectation value to $y$ in \twohyper.  This means
that the worldsheet action will contain a term of the form $\int d^2 z
V_y^{(0,0)}$, with $V_y^{(0,0)}$ given in \dvertex.  In a noncompact
model it may be possible to keep the coefficient small (in a compact
model we might expect some kind of quantization, as is the case with
\NSNS\ flux, making it difficult to treat the torsion perturbatively).
To first order in this perturbation, the cubic term in the
superpotential will be nonvanishing if the correlator
\eqn\perturbedcorr{
	A_1 = \langle V_{\phi^i}^{(-1,-1)} V_{\phi^j}^{(-1,-1)} V_{\hat{\bar{y}}^k}^{(0,0)}
		\int d^2 z V_y^{(0,0)} \rangle
}
is nonvanishing.  The form of the vertex operators indicates that
there are terms in $A_1$ corresponding to expectation values of eight
worldsheet fermions with precisely the right spacetime and worldsheet
index structure to soak up the zero modes in the
one-worldsheet-instanton sector.

Of course, this computation is at best schematic, and valid for local models.
It would be interesting and important to describe worldsheet instantons and their
effects directly in compactifications with magnetic flux and 
$SU(3)$ or $SU(3)\times SU(3)$ structures.

\newsec{Conclusions}

%\BerkovitsWR
\lref\BerkovitsWR{
  N.~Berkovits,
  ``Covariant quantization of the Green-Schwarz superstring in a Calabi-Yau
  background,''
  Nucl.\ Phys.\  B {\bf 431}, 258 (1994)
  [arXiv:hep-th/9404162].
  %%CITATION = NUPHA,B431,258;%%
}
%\AganagicPY
\lref\AganagicPY{
  M.~Aganagic, C.~Beem and S.~Kachru,
  ``Geometric Transitions and Dynamical SUSY Breaking,''
  arXiv:0709.4277 [hep-th].
  %%CITATION = ARXIV:0709.4277;%%
}
%\AganagicZM
\lref\AganagicZM{
  M.~Aganagic and C.~Beem,
  ``Geometric Transitions and D-Term SUSY Breaking,''
  arXiv:0711.0385 [hep-th].
  %%CITATION = ARXIV:0711.0385;%%
}
%\BeckerEF
\lref\BeckerEF{
  M.~Becker, K.~Dasgupta, S.~H.~Katz, A.~Knauf and R.~Tatar,
  ``Geometric transitions, flops and non-Kaehler manifolds. II,''
  Nucl.\ Phys.\  B {\bf 738}, 124 (2006)
  [arXiv:hep-th/0511099].
  %%CITATION = NUPHA,B738,124;%%
}
%\BeckerQH
\lref\BeckerQH{
  M.~Becker, K.~Dasgupta, A.~Knauf and R.~Tatar,
  ``Geometric transitions, flops and non-Kaehler manifolds. I,''
  Nucl.\ Phys.\  B {\bf 702}, 207 (2004)
  [arXiv:hep-th/0403288].
  %%CITATION = NUPHA,B702,207;%%
}
%\BeckerSH
\lref\BeckerSH{
  K.~Becker, M.~Becker, P.~S.~Green, K.~Dasgupta and E.~Sharpe,
  ``Compactifications of heterotic strings on non-Kaehler complex  manifolds.
  II,''
  Nucl.\ Phys.\  B {\bf 678}, 19 (2004)
  [arXiv:hep-th/0310058].
  %%CITATION = NUPHA,B678,19;%%
}
%\AdamsKB
\lref\AdamsKB{
  A.~Adams, M.~Ernebjerg and J.~M.~Lapan,
  ``Linear models for flux vacua,''
  arXiv:hep-th/0611084.
  %%CITATION = HEP-TH/0611084;%%
}
%\DermisekQI
\lref\DermisekQI{
  R.~Dermisek, H.~Verlinde and L.~T.~Wang,
  ``Hypercharged Anomaly Mediation,''
  arXiv:0711.3211 [hep-ph].
  %%CITATION = ARXIV:0711.3211;%%
}
%\VerlindeQK
\lref\VerlindeQK{
  H.~Verlinde, L.~T.~Wang, M.~Wijnholt and I.~Yavin,
  ``A Higher Form (of) Mediation,''
  arXiv:0711.3214 [hep-th].
  %%CITATION = ARXIV:0711.3214;%%
}

A principle lesson of this paper is that the string worldsheet
provides a powerful organizing principle for mathematical structures
that describe $\CN=1$ type II string backgrounds with intrinsic
torsion.  Furthermore, it is a {\it necessary} organizing tool, since
generically such backgrounds will have string-scale features such as
nongeometric fluxes, and physical quantities will have contributions
that are nonperturbative in $\alpha'$.

However, an additional caveat is that in designing realistic string
compactifications, one typically includes Ramond-Ramond fields.  In
this case, one might have to appeal to a formalism such as the one
described in \refs{\BerkovitsWR}.

To our minds, further progress in these directions require the
construction of more explicit examples of string compactifications
with these features.  Such features can arise either classically or be
sourced by quantum effects.  Some progress on classically stabilized
moduli, with auxiliary fields for both hypermultiplets and vector
multiplets, has already been made for type IIA
vacua \refs{\DeWolfeUU}.  Another possibility would be to $N=(2,1)$
gauged linear sigma models describing backgrounds with flux and
torsion, taking the recent, elegant construction \refs{\AdamsKB}\ for
$(0,2)$ models as a starting point.

To see how quantum effects might generate the features described in
this paper, we note that superpotentials generated by open string
gauge theory effects or by D-instantons will generically depend on the
K\"ahler moduli and on the dilaton. If supersymmetry is broken by such
F-terms, the auxiliary fields we have described here should be sourced
by quantum effects, or should appear classically as duals via a
geometric transition as in \refs{\AganagicPY,\AganagicZM}. In particular,
for the scenario described in \refs{\DermisekQI,\VerlindeQK}, the 
bulk fields mediating supersymmetry breaking are type IIB RR axions, and
we expect that the corresponding F-terms will be of the type described in the paper
(or at worst will descend from auxiliary fields in another off-shell description of the
hypermultiplets).

Another direction of work which should be relatively straightforward
is the extension to heterotic flux
compactifications \refs{\BeckerYV\BeckerSH
\BeckerQH\AlexanderEQ\BeckerII-\BeckerEF}. In these cases
there are already a small set of interesting examples.

\vskip 1cm

\centerline{\bf Acknowledgements}

We would like to thank Ruben Minasian for early collaboration on this
project, and for many useful discussions. We would also like to thank
Allan Adams, Oliver DeWolfe, Mariana Gra\~na, Shamit Kachru, John
McGreevy, Howard Schnitzer, Alessandro Tomasiello and Daniel Waldram
for helpful discussions.  Part of this project was carried out while
two of us (A.L. and M.S.) were attending the KITP program in string
phenomenology in the fall of 2006.  A.L. and B.W. would like to thank
the theory group at CEA Saclay for their hospitality during part of
this project. A.L. would also like to thank the MIT Center for
Theoretical Physics for their generous hospitality at various times
over the course of this project.  This research was supported in part
by the National Science Foundation under Grant
No. PHY99-07949. A.L. and T.S. are supported by NSF grant PHY-0331516,
by DOE Grant No.~DE-FG02-92ER40706, and by a DOE Outstanding Junior
Investigator award.  M.B.S. is supported in part by DOE grant
DE-FG02-95ER40893.  B.W. is supported in part by NSF grant
PHY-00-96515, and by the Frank and Peggy Taplin Membership at the Institute for Advanced Study.

\appendix{A}{Mathematical conventions}

The $SU(3)$ and $SU(3)\times SU(3)$ structures employed in this paper
can be defined, equivalently, in terms of spinors $\eta^N_{\pm}$ or in
terms of the differential forms $J^N$ and $\Omega^N$, for $N=1,2$. The
relationship between the two is given in
Eqs.~\suthreeacs,\suthreetopform\ and \JsandOmegas.  The conventions
given below ensure the consistency of these definitions, and are used
to compute the auxiliary fields in terms of the intrinsic torsion
classes.

\subsec{Spinor conventions}

The gamma matrices $\gamma^A$, for $A = 1,\ldots, 6$ are $8\times 8$
complex matrices representing the Clifford algebra
$\{\gamma^A,\gamma^B\} = 2 \eta^{AB}$ where $\eta^{AB}$ is the flat
metric in the vielbein basis.  The gamma matrices with spacetime
indices are $\gamma_m = e_m^A \gamma_A$, where $e_m^A$ is the vielbein
for the six-dimensional Euclidean compactification manifolds $M$. The gamma
matrices
\eqn\prodgamma{\gamma_{A_1\ldots A_k}
= \gamma_{[ A_1} \gamma_{A_2} \ldots \gamma_{A_k]}, }
including the identity matrix (for $k=0$) and the chirality operator
\eqn\chiralityop{
	\gamma_7 = i \gamma_{123456}, }
(for $k=6$) are all linearly independent.  Note that for the Euclidean
space $M$ we define $\gamma_7$ in \chiralityop\ with a factor of $i$
so that $(\gamma_7)^2 = 1$.

The chiral spinors $\eta_{\pm}$ satisfy the conditions
\eqn\chiralspin{
\eqalign{ 
\half \left(1 \pm \gamma_7\right) \eta_{\pm} & = \eta_{\pm}, \cr
\half \left(1 \mp \gamma_7\right) \eta_{\pm} & = 0,
}}
as well as the Fierz identity
\eqn\fierzident{
	\eta_{\pm} \otimes \eta^{\dagger}_+ = 
	\frac{1}{8} \sum_{k = 0}^6 \frac{1}{k!} 
        \eta^{\dagger}_+ \gamma_{A_1\ldots A_k}
		\eta_{\pm} \gamma^{A_k \ldots A_1,}
}
where the gamma matrix for the $k = 0$ term is the identity matrix.
Note that this differs by a factor of 2 from the conventions in
Refs.~\refs{\GranaBG,\GranaNY}. The factor of $1/8$ is fixed by taking
the trace of both sides; the trace over the identity matrix on the
right hand side gives a factor of $8$ since the gamma matrices are
$8\times 8$ matrices if they are to act on spinors of both
chiralities.\foot{We would like to thank A. Tomasiello for explaining
this, and for his patient and generous help with sorting out the
conventions.}

It is also useful to define gamma matrices with complex indices, since
we use these extensively in this work. As usual, one can define
$\gamma^i$ and $\gamma^{\bar\imath}$ to have the anticommutators
\eqn\holgam{
   \{ \gamma^i , \gamma^{\jb} \} = 2g^{i \jb} 
   \qquad {\rm and} \qquad 
   \{ \gamma^i , \gamma^j \} = \{ \gamma^{\bar\imath}, \gamma^{\jb} \} = 0.}
These matrices act as fermionic raising and lowering operators, and
can be used to build the spinor representations of $\Spin(6)$.  As a
consequence of Eqs.~\JsandOmegas, the spinors $\eta^N_{\pm}$ of
Eq.~\mwdecompose\ satisfy $\gamma^{\bar\imath} \eta^N_-
= \gamma^i \eta^N_+ =0$.

\subsec{Metrics, $p$-forms, and the almost complex structure}

Recall that the components of $p$-forms are defined as:
\eqn\pformnorm{
\eqalign{
	A & = \frac{1}{p!} A_{n_1\ldots n_p} dx^{n_1} \wedge \ldots \wedge dx^{n_p}\cr
	& = \frac{1}{p!} A_{B_1\ldots B_p} e^{B_1} \wedge \ldots \wedge e^{B_p}
}}
where $e^B = e^B_m dx^m$.

As in \S3.2, it is useful to define a complex vielbein $\{e^a,\e^{\bar
a}\}$ (cf. Eq.~\viels).  Then, in the vielbein basis, the metric is
becomes
\eqn\metriccomplex{
	g_{a\bar{b}} = \half \eta_{a\bar{b}}
}
with $\eta_{1\bar{1}} = \eta_{2\bar{2}} = \eta_{3\bar{3}} = 1$ and
other components of $\eta_{a\bar b}$ vanishing.  In the same basis,
the fundamental form and canonically normalized $(3,0)$ form are
\eqn\structureagain{
\eqalign{
	J & = i g_{a\bar{b}} e^a \wedge e^{\bar{b}}, \cr
	\Omega & = \frac{1}{3!} \epsilon_{abc} e^a \wedge e^b \wedge e^c.
}}
Here, the antisymmetric symbol $\epsilon_{abc}$ is defined by
$$\eps_{123} = \eps_{231} = \eps_{312} = 1,\qquad
  \eps_{213} = \eps_{321} = \eps_{132} = -1,$$  
with $\eps_{abc}=0$ otherwise.  Similarly, we define $\epsilon^{\bar
a\bar b\bar c}$ by
$$\eps^{\bar1\bar2\bar3} = \eps^{\bar2\bar3\bar1} = \eps^{\bar3\bar1\bar2} 
   = 1,\qquad
  \eps^{\bar2\bar1\bar3} = \eps^{\bar3\bar2\bar1} = \eps^{\bar1\bar3\bar2}
  = -1,$$  
with $\eps^{\bar a\bar b\bar c}=0$ otherwise.  In terms of the latter,
\eqn\raiseomega{
\eqalign{
	\Omega^{\bar{a} \bar{b} \bar{c}} & = g^{a\bar{a}}g^{b\bar{b}}g^{c\bar{c}} \Omega_{abc} \cr
	& = g^{a\bar{a}}g^{b\bar{b}}g^{c\bar{c}} \eps_{abc} \cr
	& = 8 \eps^{\bar{a}\bar{b}\bar{c}}
}}

For the three-form field strength $H$, we follow the conventions of
Polchinski \refs{\PolchinskiRQ,\PolchinskiRR}:
\eqn\hfromB{
	H_{mnp} = \p_m B_{np} + \p_n B_{pm} + \p_p B_{mn},
}
which is equivalent to $H = d B$, with the forms normalized in terms
of their components as in \pformnorm.
\listrefs

\end